\documentclass[aps,pra,amsmath,amssymb,11pt,final,tightenlines,twoside,onecolumn,nofloats,nofootinbib,superscriptaddress,
showkeys,showkeywords]{revtex4-2}

\usepackage[T2A]{fontenc}
\usepackage[utf8x]{inputenc}
\usepackage[russian,english]{babel}
\usepackage{graphicx}

\begin{document}

\selectlanguage{english}

\noindent {\it ASTRONOMY REPORTS, 2025, volume , № }
\bigskip\bigskip  \hrule\smallskip\hrule
\vspace{35mm}


\title{OLD OPEN CLUSTERS NGC~188 AND M 67 IN GAIA DR3}

\author{\bf\copyright~2025 \firstname{A.~F.}~\surname{Seleznev}}
\email{anton.seleznev@urfu.ru}
\affiliation{Ural Federal University, Ekaterinburg, Russia}

\author{\bf\firstname{M.~V.}~\surname{Kulesh}}
\affiliation{Ural Federal University, Ekaterinburg, Russia}

\begin{abstract}
\vspace{3mm}
\received{12.06.24}
\revised{12.06.24}
\accepted{12.06.24} 
\vspace{3mm}

We performed a statistical study of two old open clusters NGC 188 and M 67 using Gaia DR3 data.
No tidal tails of the clusters were detected, which most likely had been destroyed when the cluster passed through the Galactic plane.
The size estimates of the clusters depend on the range of astrometric parameters and stellar magnitudes of the stars used for star counts.
The mass spectra of two clusters differ significantly.
NGC 188 shows a deficit of low-mass stars compared to M 67.
In the halo region of NGC 188 (compared to the core region of the cluster), there is a relative excess of the low-mass stars (just as in the case of M 67) and a deficit of stars in the mass range from 0.66 to 0.9 solar masses.
Comparison with the Hunt\&Reffert samples showed that almost all the stars from these samples are contained among the stars we selected for counting.
Moreover, the group probability of these stars belonging to clusters, estimated by the uniform background method, is higher than 60\%.
It is shown that the velocity dispersion of single stars (selected according to the `stellar magnitude -- color index' diagrams) is significantly smaller than the velocity dispersion of unresolved binary stars.

\end{abstract}

\maketitle

\section{INTRODUCTION}

There are two approaches to the study of star clusters, which we call as detailed and statistical ones \cite{Danilov2020}.
In a detailed approach, one determines the cluster membership probabilities of individual stars and compiles the samples of probable cluster members on this base.
The cluster membership probabilities are estimated from the positions of stars in multidimensional spaces of astrometric parameters of stars (sometimes, the photometric parameters are added).
The detailed approach became possible and successfully progresssed thanks to the work of the Gaia space mission \cite{Gaia}, and its Gaia DR2 \cite{GaiaDR2} and Gaia DR3 \cite{GaiaDR3} catalogues.
The best known catalogues of probable cluster members are those by Cantat-Gaudin et al. \cite{Cantat-Gaudin2024} and Hunt \& Reffert \cite{H&R2023,H&R2024}.
Samples of probable cluster members obtained in this way can be used, for example, to study the kinematics and dynamics of clusters \cite{Danilov2020,Danilov2021a,Danilov2021b}.

However, one cannot obtain complete samples of probable cluster members with a detailed approach.
The reason for this is that a fraction of the stars (approximately 19\%) in Gaia DR3 have only two-parameter solutions.
In addition, the Gaia DR3 astrometric solutions are obtained based on a single star motion model, so many binary and multiple stars have large values of the so-called renormalized unit weight error (RUWE>1.4) and/or large relative errors of astrometric parameters.
Such stars are usually not included in samples of probable cluster members.
The work \cite{Tagaev3532} shows that in the case of the cluster NGC 3532, up to a half of all the cluster stars could have been lost during the selection of probable members.
It is important that this will result in the loss of a large number of unresolved binary and multiple stars, which could negatively impact the search for cluster members with unusual properties (blue stragglers, cataclysmic variable stars, etc.).
We note that NGC 3532 is located very close to the galactic equator (its galactic latitude is 1.38 degrees), and it is projected onto a rich stellar background.
For NGC 188 and M 67, located at galactic latitudes of 22.37 and 31.92 degrees, respectively, this effect is not noticeable \cite{Tagaev_stat}.

In the statistical approach, we compare the region containing the cluster being studied with one or more comparison regions containing only field stars.
The goal of the statistical approach is usually to obtain the distribution functions that characterize the cluster: the surface number density function, the brightness function and the luminosity function, the mass spectrum or the mass function.
The main obstacle in this case is the large-scale fluctuations in the density of field stars, which can mask the cluster.

To reduce the influence of fluctuations, we perform a preliminary selection of stars based on astrometric parameters (parallaxes and proper motions).
As a result, the completeness of the resulting sample is also vanished (at least due to stars with two-parameter solutions of Gaia DR3).
Usually, we do not take into account the RUWE parameter.
However, we can lost some of the cluster members with astrometric parameters differing significantly from the average values due to errors.
However, one can hope that due to the less rigorous selection of stars in the statistical approach, fewer stars will be lost than in the detailed one (this is confirmed in the present work).
A statistical approach can also yield a sample of probable cluster members \cite{Danilov2020}, but it will certainly contain field stars.
The uniform field method \cite{Danilov2020} allows us to estimate the group cluster membership probability of these stars.

The distribution functions obtained by the statistical approach provide a more complete picture of the properties of the cluster than the same functions obtained using the detailed approach due to smaller losses of stars.
Moreover, the implementation of the statistical approach is significantly simpler than the methods of analyzing the multidimensional spaces when obtaining the cluster membership probability of individual stars.

We selected for our study two old, well-studied open star clusters (OSCs), NGC 188 and M 67 (NGC 2682).
One of the reasons for the interest in these clusters is the results on the distribution of stars of different masses in them, obtained almost 50 years ago \cite{Tinsley&King1976,McClure&Twarog1977}.
According to these results, red giants (RGs) in these clusters are distributed in a larger volume than the main sequence (MS) stars.
Similar results for six other old open clusters were obtained in \cite{Hawarden1975}.
Authors of \cite{McClure&Twarog1977,Hawarden1975} proposed a hypothesis that the RGs have already experienced a significant loss of mass and have been redistributed themselves in space due to collisional relaxation to explain this fact.
In the case of M 67, the authors of \cite{Tinsley&King1976} believed that mass loss by stars cannot explain the observed RGs distribution, and wrote that brighter giants do not show a relaxed distribution at all.
According to modern PARSEC isochrones \cite{PARSEC}, the masses of the observed stars above the MS turn-off point have masses similar to, or even slightly greater than the stars in the upper MS.
Therefore, the hypothesis about the mass loss by RG stars does not work.
In addition, the results of \cite{McClure&Twarog1977, Tinsley&King1976} apply to the very central part of the clusters: in the case of NGC 188, approximately for a region of 20x30 arcmin, and for M 67 --- for a region with a radius of approximately 40 arcmin.
According to the data of \cite{H&R2023,H&R2024}, the outermost radius of NGC 188 is 1.16 degrees, and the outermost radius of M 67 is 1.68 degrees.
Also, the results of \cite{McClure&Twarog1977, Tinsley&King1976} can be explained by the incompleteness of the data on faint stars in the outer part of the fields studied by these authors.
On the other hand, if the results of \cite{McClure&Twarog1977, Tinsley&King1976, Hawarden1975} are confirmed, such a distribution of RGs can be explained by the non-stationarity of clusters with respect to the mean gravitational field.
In this regard, we decided to perform a statistical study of these clusters at the base of modern Gaia DR3 data.

\begin{table}
\caption{Clusters' main parameters from \cite{Dias+2021}}
\label{tab:cluster_par}
\begin{tabular}{c|c|c|c|c|c|c|c|c|c}
\hline
Cluster &    l    &    b   &  $\mu_\alpha$  &  $\mu_\delta$  &   $\varpi$    &     d       &   $\log t$ &  $[Fe/H]$ &  $A_V$  \\
\hline
          &  deg.  & deg.  &    mas/yr     &   mas/yr      &      mas      &     pc      &          &    & \\
\hline

NGC 188   & 122.839 & 22.370 & -2.30$\pm$0.18 & -0.96$\pm$0.17 & 0.51$\pm$0.05 & 1860$\pm$40 & 9.79$\pm$0.03 & 0.11$\pm$0.02 & 0.23$\pm$0.03 \\
M 67      & 215.692 & 31.922 &-11.00$\pm$0.22 & -2.96$\pm$0.22 & 1.14$\pm$0.05 &  865$\pm$18 & 9.58$\pm$0.03 & 0.07$\pm$0.05 & 0.13$\pm$0.04 \\

\hline
\end{tabular}
\end{table}

Both NGC 188 and M 67 have a very rich history of exploration.
These investigations concerned generally the study of variable stars, the chemical composition of stars, obtaining the fundamental characteristics of clusters (distance, age, reddening) based on photometry data, and research into the population of binary stars and blue stragglers.
NGC 188 is one of the key objects of the WOCS project (WIYN Open Cluster Study, WIYN means a consortium of the Universities of Wisconsin, Illinois, and Yale and the US National Astronomical Observatory).
During the course of this project, the very important data on spectroscopic binaries were obtained at the base of the long-standing monitoring observations.
In particular, authors of \cite{188_Geller_Mathieu} showed that at least 80\% of the blue straggler stars are components of spectroscopic binaries.

Statistical studies of NGC 188 and M 67 included an obtaining of the brightness function (luminosity function) and the mass function (mass spectrum).
However, almost all of these studies were limited to the central part of the clusters.
In these studies a comparison area (if such was used) was chosen as being projected onto the cluster (according to modern data of \cite{H&R2023,H&R2024}).
For NGC 188 these are the works \cite{Caputo+1990,vonHippel_Sarajedini_1998AJ,Sarajedini+1999AJ,Bonatto+2005A&A,Elsanhoury+2016NewA}.
Sometimes studies have produced opposite results.
Thus, in \cite{Caputo+1990,vonHippel_Sarajedini_1998AJ,Sarajedini+1999AJ} authors noted a decrease in the luminosity function of NGC 188 towards faint stars, while authors of \cite{Elsanhoury+2016NewA} noted its growth.
For M 67, the luminosity function was plotted in the works \cite{van_den_Bergh1957AJ,Kholopov_Artyukhina_1965SvA,Bonatto_Bica_2003A&A,Davenport_Sandquist_2010ApJ}.
In the works \cite{Kholopov_Artyukhina_1965SvA, Davenport_Sandquist_2010ApJ} the authors advanced, respectively, to $r=64.7$ and $r=60$ arcmin from the cluster center, but this is also noticeably smaller than the size of M 67 according to the modern data \cite{H&R2023,H&R2024}.
In these studies, a deficit of low-mass stars was discovered in the region of the cluster core ($r<7.5$ arcminutes in \cite{Kholopov_Artyukhina_1965SvA} and $r<8.24$ arcminutes in \cite{Davenport_Sandquist_2010ApJ}).
In addition, the authors of \cite{Davenport_Sandquist_2010ApJ} noted the elongation of the cluster halo approximately in the direction of the proper motion vector.

Some works performed the N-body numerical modeling of NGC 188 and M 67.
We should mention here the work of Chumak et al. \cite{Chumak+2010MNRAS}.
Authors of \cite{Chumak+2010MNRAS} studied the formation of tidal tails of NGC 188 (the authors tried to detect them using the 2MASS data, but failed).
In the paper \cite{Hurley+2005MNRAS} the authors investigated the change in the cluster luminosity function over time.

We list the main characteristics of these two clusters, based on data from the Dias et al. \cite{Dias+2021} catalogue, in Table \ref{tab:cluster_par}.
Table \ref{tab:cluster_par} contains in order galactic longitude and galactic latitude in degrees, proper motions in right ascension and declination in milliarcseconds per year, parallax in milliarcseconds, heliocentric distance in parsecs, logarithm of age (age in years), metallicity and total extinction in the V band in stellar magnitudes.

The aim of this work is to conduct a statistical study of NGC 188 and M 67 (NGC 2682) based on Gaia DR3 data \cite{GaiaDR3}.
Namely, we plot the surface density maps, determine the sizes of clusters, obtain brightness functions and mass spectra, investigate the distribution of stars of different masses, and compare our results with the results of \cite{H&R2023,H&R2024} obtained within the framework of a detailed approach.
Section II describes the acquisition of initial data, preliminary selection of stars, and standard data processing.
Section III presents the results of the study of the distribution of the surface density.
Section IV is devoted to obtaining the brightness functions and mass spectra.
In Section V we compare our results with the Hunt \& Reffert sample \cite{H&R2023} and analyse the cumulative stellar density distributions.
In Section VI, the velocity dispersion is calculated for single and unresolved binary cluster stars.
We discuss the main results of the work in Section VII.

\begin{figure}[h]
\includegraphics[width=0.48\textwidth]{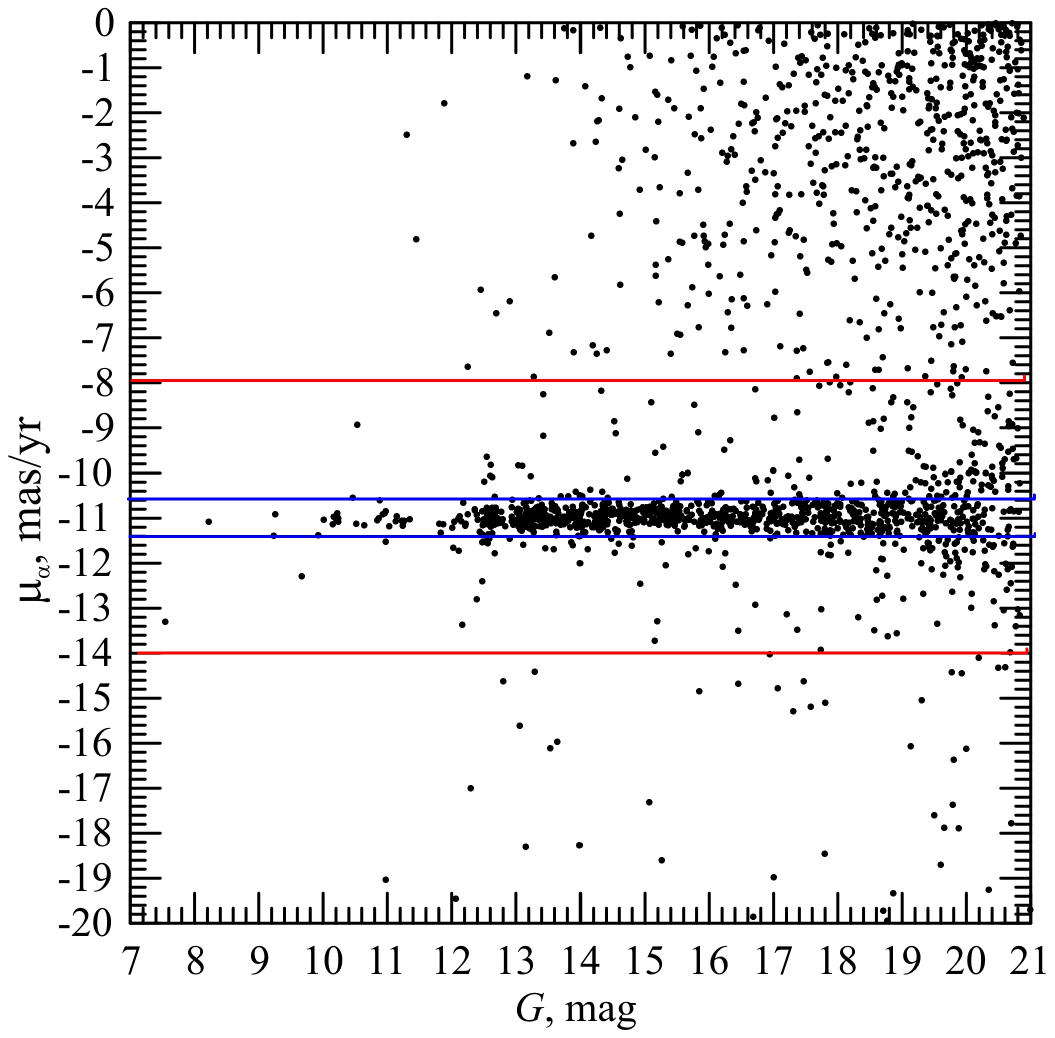}
\caption{Diagram `proper motion in right ascension $\mu_\alpha$ vs. G magnitude' for stars within 30 arcmin from the center of the M 67 cluster. Red lines show the `standard' selection range, blue lines show the `strict' selection range.}
\label{pmra-G_67}
\end{figure}

\begin{figure}[h]
\includegraphics[width=0.48\textwidth]{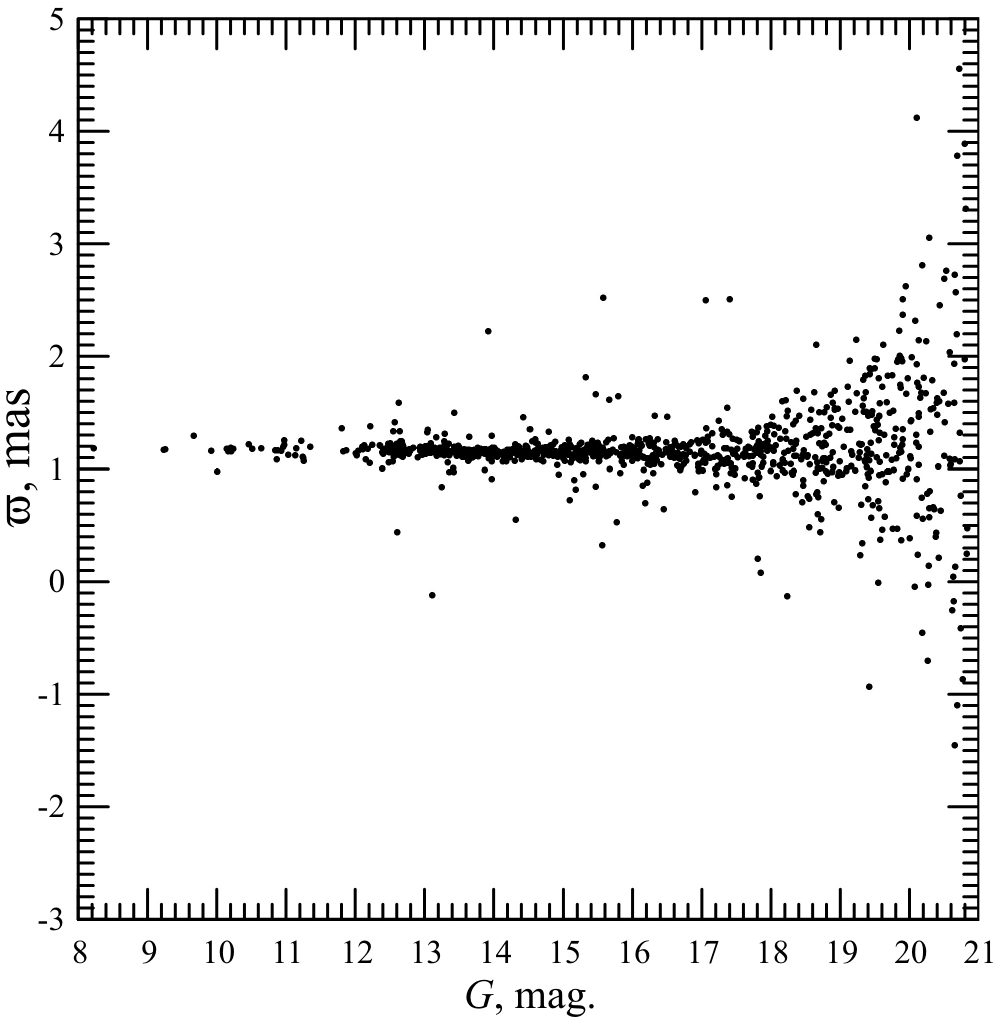}
\caption{Diagram `parallax $\varpi$ vs. G magnitude' for stars within 30 arcmin from the center of the M 67 cluster.}
\label{plx-G_67}
\end{figure}

\section{Initial data and their standard processing}

In order to select the ranges of the astrometric parameters confined stars for cluster studies, we first downloaded data from the Gaia DR3 catalogue (using the VIZIER system) within circles of a 20-arcmin radius for NGC 188 and 30-arcmin radius for M 67, centered according to the \cite{Dias+2021} catalogue (this corresponds to a region with a radius of 10.8 pc for NGC 188 and 7.5 pc for M 67).
We determined the range boundaries using the `astrometric parameter -- stellar magnitude $G$' diagrams.

First of all, we select a diagram in which the cluster stars (forming a horizontal stripe on such a diagram) are best separated from the field stars.
For NGC 188 this is the `$\mu_\delta$ -- $G$' diagram, and for M 67 it is the `$\mu_\alpha$ -- $G$' diagram (Fig. \ref{pmra-G_67}).
Fig. \ref{pmra-G_67} clearly shows the expansion of the cluster band at $G>19$ mag.
We chose the range for the `standard' selection of stars in such a way that this range includes all the stars of the extended cluster band at $G=21$ mag.
This range is shown in Fig. \ref{pmra-G_67} by red lines.
The width of this range corresponds to $\pm12.3$ km/s relative to the mean cluster velocity in the direction of right ascension.
It is much larger than the possible velocity dispersion of the cluster stars.
The blue lines show the range for `strict' selection.
Its width corresponds to $\pm1.64$ km/s.
This value is also greater than the possible velocity dispersion in one direction, but due to errors in proper motion, some of the cluster stars will be lost with the strict selection.

At the next step, we select the ranges of proper motion along the second coordinate, and after that, ranges of parallax.
In the case of M 67, it turned out that we do not need to set the range of parallax, since after applying the standard ranges on proper motions, there were nearly no field stars left on the `parallax -- stellar magnitude' diagram (Fig. \ref{plx-G_67}).
In Fig. \ref{plx-G_67} it is evident that the cluster strip begins to expand already at approximately $G>17$ mag.

In the case of NGC 188, we should establish the parallax range, since the NGC 188 cluster strip on the `astrometric parameter -- stellar magnitude $G$' diagrams is more polluted by the field stars.

As a result, we obtained the following ranges for parallaxes and proper motions.
`Standard' ranges for NGC 188:

\begin{equation}
\begin{array}{lll}
\varpi\in[0;1.3] \; \mbox{mas}  \\
\mu_\alpha\in[-3.3;-1.3]  \; \mbox{mas/yr} \\
\mu_\delta\in[-1.8;-0.2]  \; \mbox{mas/yr} .
\end{array}
\label{interval_1_188}
\end{equation}

\noindent `Strict' ranges for NGC 188:
\begin{equation}
\begin{array}{lll}
\varpi\in[0.3;0.8] \; \mbox{mas}  \\
\mu_\alpha\in[-2.7;-1.9]  \; \mbox{mas/yr} \\
\mu_\delta\in[-1.4;-0.6]  \; \mbox{mas/yr} .
\end{array}
\label{interval_2_188}
\end{equation}

\noindent `Standard' ranges for M 67:
\begin{equation}
\begin{array}{ll}
\mu_\alpha\in[-14;-8]  \; \mbox{mas/yr} \\
\mu_\delta\in[-6;0]  \; \mbox{mas/yr} .
\end{array}
\label{interval_1_67}
\end{equation}

\noindent `Strict' ranges for M 67:
\begin{equation}
\begin{array}{ll}
\mu_\alpha\in[-11.4;-10.6]  \; \mbox{mas/yr} \\
\mu_\delta\in[-3.4;-2.6]  \; \mbox{mas/yr} .
\end{array}
\label{interval_2_67}
\end{equation}

We do not use the radial velocity data for the selection of stars.
Unfortunately, the data on radial velocities are very fragmentary, despite a significant progress in this direction.
In addition, with the statistical approach to cluster study, we do not need a very strict selection of stars (as in the case of the detailed approach).
Moreover, the statistical approach requires the presence of field stars in the studied sample.
We select stars by proper motions and parallaxes only to reduce the large-scale fluctuations of the field star density.

For further study, we selected stars from the Gaia DR3 catalogue from a region with a radius of 510 arcmin around the cluster centers, satisfying the constraints (\ref{interval_1_188})-(\ref{interval_2_67}).

After this, we refine the coordinates of the cluster centers.
For this purpose, we plotted the linear density distributions by galactic coordinates using the Kernel Density Estimator (KDE) method \cite{Silverman1986}.
Using the galactic coordinate system when working with open clusters is preferable to using the equatorial system, since the galactic latitude values of clusters are usually small.
This helps to reduce the influence of the meridian convergence.
To plot the linear density distributions, a biquadratic kernel \cite{Silverman1986} with a half-width of 0.16 degrees was used.
For NGC 188, we obtained the center coordinates of $l_c=122.850$ and $b_c=22.375$, very close to the values from \cite{Dias+2021}.
For M 67, the obtained coordinates of the cluster center do not differ from the values of \cite{Dias+2021} within the linear density distribution step (0.005 degrees or 18 arcsec).
Therefore, we used the center coordinates of M 67 from Table \ref{tab:cluster_par}.

The final procedure of standard processing of the initial data is an obtaining the rectangular coordinates in the tangential plane (touching the celestial sphere at the point of the cluster center).
We used the equidistant azimuthal (polar) projection for this purpose.
The advantage of this method of projecting a sphere onto a plane is that the distance of the point (star) from the pole (center of the cluster) does not change.
Formulas for transition to a rectangular coordinate system are:

\begin{equation}
\begin{array}{ccccc}
\cos \rho=\sin b\sin b_c+\cos b\cos b_c\cos(l_c-l) \\
\sin \varphi=-\frac{\displaystyle\sin(l_c-l)\cos b}{\displaystyle\sin\rho} \\
\cos \varphi=\frac{\displaystyle\sin b-\sin b_c\cos\rho}{\displaystyle\cos b_c\sin\rho} \\
x=-\rho\sin\varphi \\
y=\rho\cos\varphi
\end{array}
\label{rectan_1}
\end{equation}
In these formulas $\rho$ is the angular distance from the center, $\varphi$ is the position angle, $l$ and $b$ are the galactic longitude and latitude, $l_c$ and $b_c$ are the galactic longitude and latitude of the center, $x$ and $y$ are rectangular coordinates.
$\varphi=0$ corresponds to the direction to the North Pole of the Galaxy, $\varphi$ increases in the counterclockwise direction as seen from the observer.
The transition to a rectangular coordinates allows us to plot the distribution of a surface density on a uniform grid.
It is convenient to make the graphical output again in the galactic coordinates.
For this, we use formulas of the inverse transformation:

\begin{equation}
\begin{array}{ccccc}
\rho^2=x^2+y^2 \\
\sin \varphi=\frac{\displaystyle x}{\displaystyle \rho} \\
\cos \varphi=\frac{\displaystyle y}{\displaystyle \rho} \\
\sin b=\sin b_c\cos\rho+\cos b_c\sin\rho\cos\varphi \\
\sin(l_c-l)=\frac{\displaystyle \sin\varphi\sin\rho}{\displaystyle \cos b}
\end{array}
\label{rectan_2}
\end{equation}

The notations in these formulas are the same as in (\ref{rectan_1}).
Using these formulas, we convert the rectangular coordinates of the grid nodes at which the density values are calculated into galactic coordinates.
This completes the standard data processing.

\section{Surface density}

First of all, we tried to detect the tidal tails of the clusters.
For this purpose, we plotted the maps of the surface density for different star selection options.
Using this approach, the tidal structures were discovered in the $\alpha$ Per cluster \cite{Nikiforova2020}.
Also, with a simple selection of probable members based on proper motions and radial velocities, the tidal tails were discovered in the old cluster Rup 147 \cite{Yeh2019}.

In addition to the selection options mentioned above (\ref{interval_1_188})-(\ref{interval_2_67}), we plotted a surface density map for all stars from a circle with a radius of 510 arcmin for M 67, as well as the maps with the additional selection of stars based on the color-magnitude diagram (CMD).
No tidal tails were detected.
This can be explained because NGC 188 and M 67 crossed the plane of the Galaxy quite a few times during their lifetime, moving along the box orbits.
The dynamic shocks that these clusters experienced as they crossed the plane of the Galaxy could have led to the destruction of the tidal tails.

We should note that authors of some works have reported the discovery of tidal tails in NGC 188 and M 67.
Gao \cite{Gao2020} used the principal component analysis to select likely cluster members and found the tidal tails around M 67 extending to a distance of 2.5 degrees from the cluster center.
The maximum value of the M 67 cluster radius obtained in our work is 133$\pm$5 arcmin or 2.2 degrees (see below in this section and Table \ref{tab:cluster_rad}).
Thus, the tidal tails from \cite{Gao2020} are located almost entirely inside M 67 according to our data (see Fig. 2 of \cite{Gao2020}).
Unfortunately, the authors of \cite{Gao2020} do not provide a description of the principal components C1, C2 and C3 they used and the method to construct them.

Kos \cite{Kos2024} reports the detection of tidal tails out to distances of 1.6 kpc for NGC 188 and 2.0 kpc for M 67.
To search for tidal tails, the author of \cite{Kos2024} used a very interesting method.
First, he calculated the galactic orbits of clusters backward in time, then modeled the dynamical evolution of clusters and tracked how their tidal tails were projected onto the celestial sphere.
Then, he compared the distribution of stars from Gaia DR3 with the model distribution of stars in the Galaxy and looked for stars located in the tidal tail region and satisfying the conditions on the velocity of stars relative to the center of the cluster.
Unfortunately, the method of \cite{Kos2024} is not free from drawbacks.
Firstly, the orbit calculations for NGC 188 and M 67 in \cite{Kos2024} are made backwards in time by several billion years, which cannot provide a reliable orbit for the cluster.
The reason is that the orbit calculations use a smoothed potential of the Galaxy and do not take into account the possible influence of massive objects in the Galactic disk on the cluster orbit, such as giant molecular clouds or gas-star complexes (GSCs) like the Gould Belt (the dynamic influence of the GSCs on open clusters is quite noticeable \cite{Danilov&Seleznev1995}).
Secondly, our experience shows that independently of the proper motions range we choose, we will find stars with such parameters throughout the sky.
Third, the author of \cite{Kos2024} considers only tidal evaporation of stars and does not take into account the stellar encounters when modeling the formation of tidal tails of clusters.

The authors of \cite{Alvarez-Baena+2024} found a tidal tail of the cluster NGC 188 up to a distance of more than 100 pc from the cluster center (and did not find a tail in M 67).
It is unclear why the tail is located only on one side of the cluster; there is no any sign of elongation on the other side.

\begin{figure}[h]
\includegraphics[width=1\textwidth]{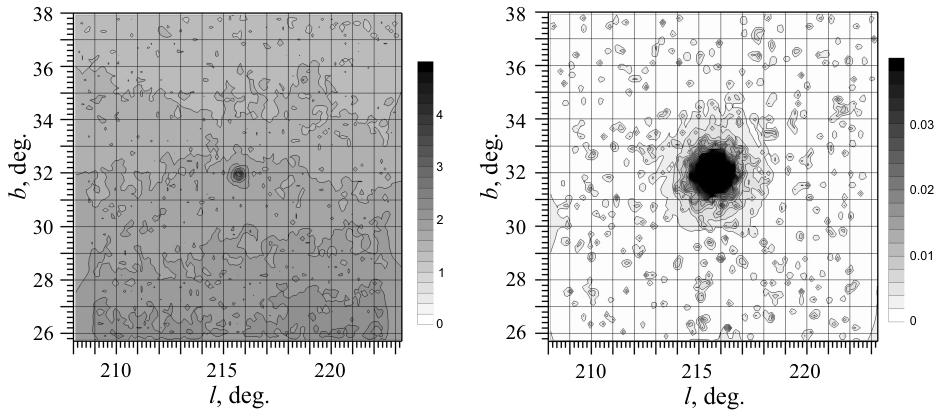}
\caption{Surface density maps for the M 67 cluster. Left: without preliminary star selection; right: after the `strict' selection (\ref{interval_2_67}).}
\label{maps_67}
\end{figure}

\begin{table}
\caption{Cluster radii derived using the method of \cite{Seleznev2016} under different stellar sample selection criteria}
\label{tab:cluster_rad}
\begin{tabular}{c|c|c|c|c}
\hline
Cluster & $G_{\rm lim}$, & Constraints on    &  $R_c$,    &  $R_c$,  \\
        & mag            & astrometric      & arcmin     &  pc      \\
        &                & parameters       &            &          \\
\hline
NGC 188 &  18             & \ref{interval_1_188} &  56$\pm$4  & 30$\pm$2 \\
        &  21             & \ref{interval_2_188} &  70$\pm$5  & 38$\pm$3 \\
\hline
M 67    &  18             & \ref{interval_1_67}  &  59$\pm$3  & 15$\pm$1 \\
        &  21             & \ref{interval_1_67}  &  94$\pm$8  & 24$\pm$2 \\
        &  21             & \ref{interval_2_67}  & 133$\pm$5  & 33$\pm$1 \\   
\hline
\end{tabular}
\end{table}

\begin{figure}[h]
\includegraphics[width=1\textwidth]{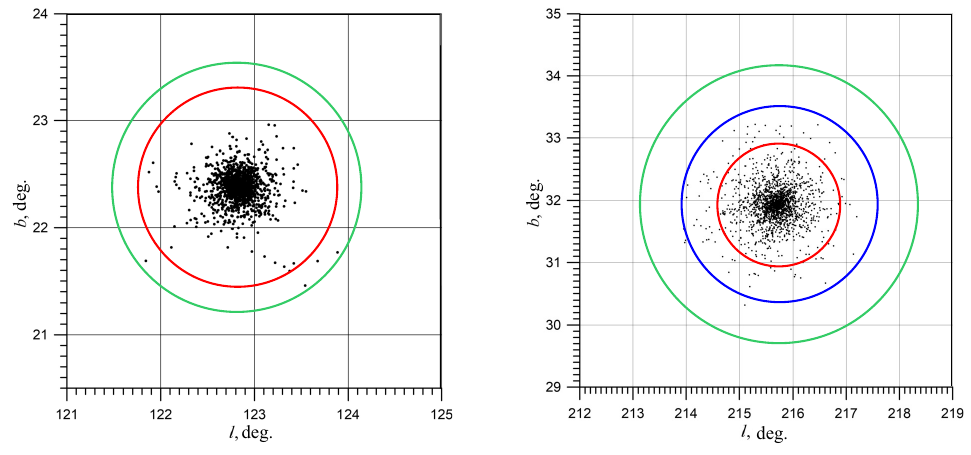}
\caption{Maps of stars from Hunt \& Reffert \cite{H&R2023} samples and cluster sizes determined from surface density profiles using the method of \cite{Seleznev2016}. Left: NGC 188. Red line --- cluster size for $G<18$ mag and constraints on the  astrometric parameters (\ref{interval_1_188}), green line --- for $G<21$ mag and constraints (\ref{interval_2_188}). Right: M 67. Red line --- $G<18$ mag and constraints (\ref{interval_1_67}), blue line --- $G<21$ mag and constraints (\ref{interval_1_67}), green line --- $G<21$ mag and constraints (\ref{interval_2_67}).}
\label{charts}
\end{figure}

Despite the failure to detect tidal tails, we obtained interesting results during the study of the surface density distribution.
Fig. \ref{maps_67} shows surface density maps of M 67 plotted using the KDE \cite{Silverman1986,KDE_OSC} for different samples of stars.
In both cases, a two-dimensional biquadratic kernel with a half-width of 10 arcminutes was used.
On the left panel of fig. \ref{maps_67}, the density map is for the sample of stars obtained without any restrictions (a circle with a radius of 510 arcmin contains 1'346'316 stars).
On the right panel --- the density map is for the sample of stars obtained with `strict' constraints on proper motions (\ref{interval_2_67}), with only 2'026 stars remaining in a circle with a radius of 510 arcmin.
In the first case, the large-scale fluctuations of the field star density prevent to detect the outer regions of the cluster with low density.
In the second case, the outer parts of the cluster are clearly visible.
Fig. \ref{maps_67} is a clear illustration of the influence of the large-scale fluctuations in the density of field stars on the results of studies of open clusters.

In addition to surface density maps, the radial surface density profiles were plotted to determine the cluster radius \cite{Seleznev2016}.
In the statistical approach, the radius of a cluster is understood as the distance from the center of the cluster at which the cluster's surface density ceases to differ from the surrounding star field.
That is, this is the distance at which the surface density of the cluster becomes smaller than the fluctuations in the density of the field stars.
This method for determining the cluster radius was proposed in \cite{Seleznev2016}, a good illustration is also given in \cite{Tagaev3532}.

To plot the radial profiles, we again used the KDE with a biquadratic kernel with a half-width of $h=10$ arcmin.
It turned out that the cluster radius changes significantly depending on the limiting magnitude of the stars and on the restrictions on the astrometric parameters imposed on the stars when constructing the sample.
We illustrate this in Fig. \ref{charts}, where the sizes of the clusters are shown as circles of different colors in comparison with the charts of the location of stars from the Hunt \& Reffert \cite{H&R2023} samples.
We list the sizes of the clusters and the conditions under which they were obtained in Table \ref{tab:cluster_rad}.
The first column of the Table \ref{tab:cluster_rad} gives the name of the cluster, the second column gives the limiting stellar magnitude $G_{lim}$, the third column gives the range of astrometric parameters, the fourth column gives the radius of the cluster in arcmin, and the fifth column gives the radius of the cluster in parsecs.
In the case of M 67, the radius increases more than 2 times and for the third sample choice it significantly exceeds the size of the cluster according to the \cite{H&R2023} data.
Thus, the statistical approach is at least as sensitive as the detailed one.

\section{Brightness function and mass spectrum}

To plot the brightness function, we used a ring with an inner radius equal to the cluster radius $R_c$ and an outer radius $R_c\sqrt{2}$ as a comparison region.
Other options for choosing the comparison area were not explored.
However, we should note that with our definition of the cluster radius (see above in the previous section), for $R>R_c$ the cluster density is not statistically different from the field one.
This makes this choice of a comparison area being quite justified.
For the comparison regions, we used the same constraints on the astrometric parameters (\ref{interval_1_188}) and (\ref{interval_1_67}) as for the cluster regions.

\begin{figure}[h]
\includegraphics[width=1\textwidth]{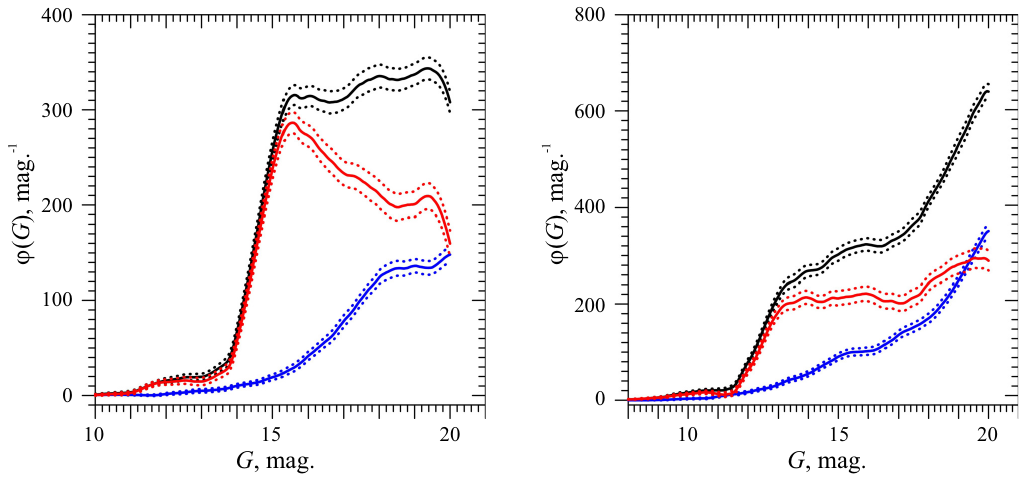}
\caption{Brightness functions. Black lines --- cluster region (cluster and field stars), blue lines --- comparison region (field stars), red lines --- difference between them (cluster stars). Solid lines — brightness function, dashed lines --- 2$\sigma$ confidence interval. Left: NGC 188, right: M 67.}
\label{lf}
\end{figure}

We accepted the values $R_c=56\pm4$ arcmin for NGC 188 and $R_c=94\pm8$ arcmin for M 67 as the cluster radius, as they most closely match the distribution of stars in the Hunt \& Reffert \cite{H&R2023} samples (see Fig. \ref{charts}).
To plot the brightness function, we used the KDE with a biquadratic kernel with a half-width of 1 mag \cite{Seleznev_LF,KDE_OSC}.
This half-width was chosen because it produced a fairly smooth function, but still retained some slight variations.
Taking into account the half-width of the kernel, the brightness functions were plotted up to the limiting magnitude $G=20$ mag in order to avoid the `undersampling'.

The obtained brightness functions are shown in Fig. \ref{lf}.
The black lines are brightness functions for the cluster region, containing cluster stars and field stars.
The blue lines are the brightness functions for the comparison region, containing only field stars.
The red lines are the difference between these functions, corresponding only to the cluster stars.
The dotted lines are the confidence interval of $2\sigma$-width plotted using the `smoothed bootstrap' method \cite{Seleznev_LF}.

Let us note, firstly, that the downward bend of the brightness function of the cluster NGC 188 is not an artifact of the method.
This is confirmed by the fact that the brightness function for the comparison region does not have such a bend.
Secondly, the brightness functions of the two clusters differ significantly.
For NGC 188, the brightness function reaches its maximum at approximately $G=15.5$ mag and then decreases almost monotonically, while for M 67 the brightness function increases to the lower brightness limit.

The conclusion about the downward bending of the brightness function of the cluster NGC 188 at $G>19$ stellar magnitudes is valid if the cluster region and the comparison region have the same completeness in this range of stellar magnitudes.
How significant can be the difference in the sample completeness for the cluster region and the comparison region?
After all, the cluster area has a higher density.
According to \cite{Fabricius+2021}, for Gaia EDR3, the sample completeness drops to 60\% for stars fainter than $G\sim19$ mag for regions with a density of $>10^5$ stars per square degree and more (for the case of globular clusters).
We estimated the density of all Gaia DR3 stars for a circle with a radius of 10 arcmin around the cluster center.
For NGC 188, the density turned out to be equal to 21,900 stars per square degree, and for M 67 --- 12,700 stars per square degree.
This is significantly less than $10^5$ stars per square degree.
This is not surprising, since both clusters are located at a fairly high galactic latitude, far from the plane of the Galaxy.

The average density of all Gaia DR3 stars in the comparison region is 10,940 stars per square degree for NGC 188 and 5,770 stars per square degree for M 67.
This is approximately 2 times less than for cluster centers.
The average density of all Gaia DR3 stars in the entire cluster region is 11,680 stars per square degree for NGC 188 and 6,090 stars per square degree for M 67.
These density values are only slightly higher than the density values in the comparison areas.
It is obvious that one of the main reasons for the incompleteness of the Gaia DR3 catalogue in the region of faint stars is the high density of stars.
The values of the density of stars in the cluster regions and in the comparison regions differ by a factor of 2 for the cluster centers and only slightly for the entire cluster region.
Thus, the conclusion about the same degree of incompleteness of samples in the cluster area and in the comparison area can be considered as quite probable.
Note that we took into account all stars from Gaia DR3 when estimating the density values, including stars with 2-parameter solutions.

We derived the brightness functions not only for each cluster as a whole, but also separately for their cores and halos.
We identified the cluster cores following the approach of P. N. Kholopov \cite{Kholopov_structure}.
We defined the core radius from the radial surface density profile as the distance between the cluster center and the end of the zone with the maximum density gradient.
Thus, we adopted $r_c = 36$ arcminutes for NGC 188 and $r_c = 68$ arcminutes for M 67.
Because we used the KDE method to construct the radial surface density profile, the core radius may be slightly overestimated, but by no more than the kernel half-width, $h = 10$ arcminutes.
We verified that reducing the core radius by up to 10 arcminutes does not noticeably affect the brightness function (only the number of halo stars changes significantly).

We then used the brightness functions of the clusters to derive their mass spectra, applying the formulas from \cite{4337}:
\begin{equation}
\psi(m)=\frac{dn}{dm} \; ,
\label{mass_spectrum}
\end{equation}
\begin{equation}
\varphi(G)=\frac{dn}{dG} \; ,
\label{lum_funct}
\end{equation}
\begin{equation}
dm=\frac{dm}{dG}\cdot dG\equiv m'_G\cdot dG \; ,
\label{mass_diff}
\end{equation}
\begin{equation}
\psi(m)=\frac{dn}{dm}=\frac{dn}{|m'_G|\cdot dG}=\frac{\varphi(G)}{|m'_G|} \; ,
\label{mass_sp_lum_fun}
\end{equation}

\noindent where $\psi(m)$ is the mass spectrum, $\varphi(G)$ is the brightness function, $m$ is the stellar mass, and $m'_G$ is the derivative of the mass–apparent magnitude relation.

\begin{figure}[h]
\includegraphics[width=1\textwidth]{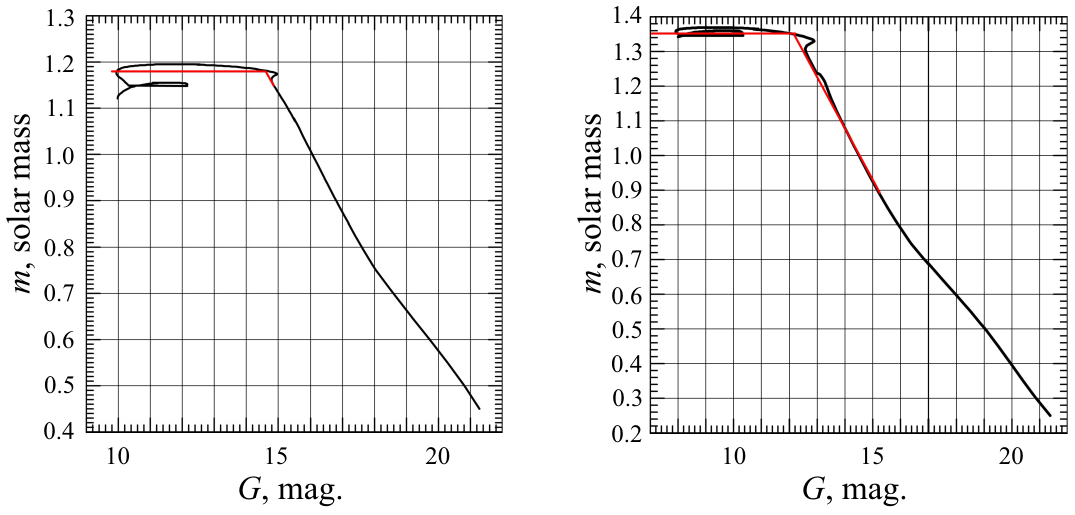}
\caption{Mass vs. apparent magnitude according to PARSEC \cite{PARSEC} isochrones (black lines). Red lines show relations used in this work. Left: NGC 188, right: M 67.}
\label{m-G}
\end{figure}

We derived the mass – apparent magnitude relation using the PARSEC theoretical isochrones \cite{PARSEC} version 1.2.
We selected the isochrone according to the cluster age and metallicity listed in Table \ref{tab:cluster_par} \cite{Dias+2021}.
For this purpose, we applied the relation $[M/H] \approx [Fe/H]$, which follows from the definition of $[M/H]$ and from equation (9) in \cite{M/H}.
We calculated the apparent magnitude $G$ and the apparent color index $BP - RP$ from the cluster distances in Table \ref{tab:cluster_par} \cite{Dias+2021}, using the relations $A_G \simeq 0.836 A_V$, $A_{BP} \simeq 1.083 A_V$, and $A_{RP} \simeq 0.634 A_V$ provided on the PARSEC isochrone output page \footnote{https://stev.oapd.inaf.it/cgi-bin/cmd\_3.8}.
These relations correspond to a G2V star and rely on the extinction curve from \cite{Cardelli+1989, O'Donnell1994} with $R_V = 3.1$.

We show the derived mass – apparent magnitude relations as black curves in Fig. \ref{m-G}.
Starting near the main-sequence turnoff point, these relations become non-unique.
To handle the high-mass region, we replaced the curves with straight-line segments (red lines in Fig. \ref{m-G}).
Thus, for stars brighter than $G = 14.6$ mag in NGC 188, we adopted a constant stellar mass of 1.18 $M_\odot$.
For M 67, we set the stellar mass to 1.35 $M_\odot$ for stars brighter than $G = 12.2$ mag.
In these regions, the derivative of the mass–apparent magnitude relation equals zero, which makes it impossible to compute the mass spectrum using equation \ref{mass_sp_lum_fun}.
We therefore derived the mass spectrum only for the intervals $G \in [14.6, 20]$ in NGC 188 and $G \in [12.2, 20]$ in M 67.
This restriction is not critical, since the primary interest lies in the mass spectrum of lower-mass stars.

To compute the derivative of the mass – apparent magnitude relation, we applied the differentiation formulas for the least-squares polynomial approximation of a function defined at equally spaced argument points \cite{Korn&Korn1968}:

\begin{equation}
y'_k\approx\frac{1}{12\Delta x}\cdot[(y_{k-2}-y_{k+2})-8(y_{k-1}-y_{k+1})] \; ,
\label{KK1}
\end{equation}
\begin{equation}
y'_k\approx\frac{1}{12\Delta x}\cdot(3y_{k+1}+10y_k-18y_{k-1}+6y_{k-2}-y_{k-3}) \; ,
\label{KK2}
\end{equation}
\begin{equation}
y'_k\approx\frac{1}{12\Delta x}\cdot(y_{k+3}-6y_{k+2}+18y_{k+1}-10y_k-3y_{k-1}) \; ,
\label{KK3}
\end{equation}
\noindent where $\Delta x$ is the step of the argument, and $y_i$ is the value of the function at the point $x_i$.
These formulas do not allow one to compute the derivative at the boundary points of the argument.
In such cases, we assigned the derivative at the nearest interior point to the boundary point.
This approach is acceptable here, since the mass–apparent magnitude relation remains nearly linear within the relevant interval (Fig. \ref{m-G}).

\begin{figure}[h]
\includegraphics[width=1\textwidth]{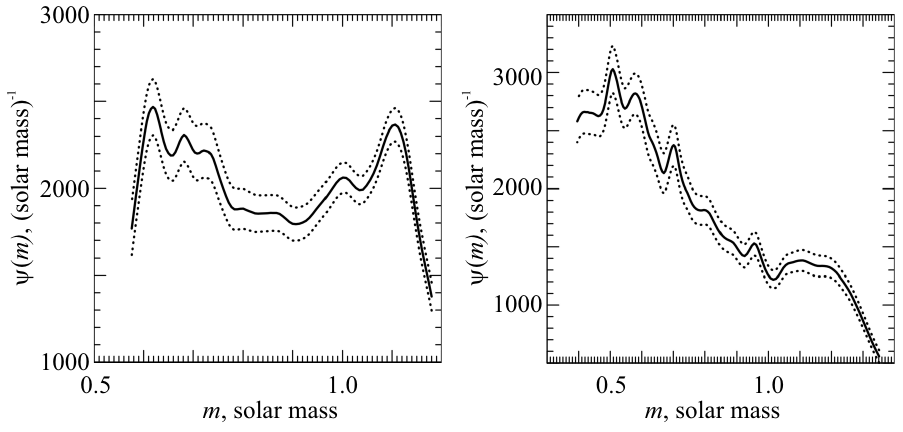}
\caption{Mass spectrum (solid line). Dashed line --- 2$\sigma$ confidence interval. Left: NGC 188, right: M 67.}
\label{mf}
\end{figure}

\begin{figure}[h]
\includegraphics[width=1\textwidth]{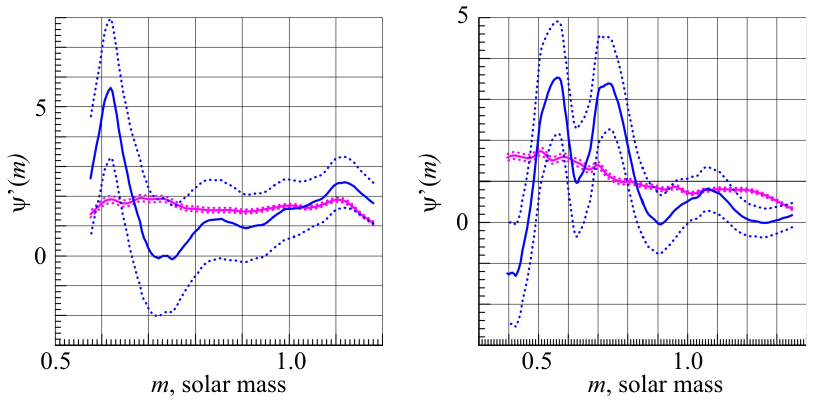}
\caption{Comparison of normalized mass functions for cluster cores (pink lines) and halos (blue lines). Solid line --- mass function, dashed line --- 2$\sigma$ confidence interval. Left: NGC 188, right: M 67.}
\label{mfnorm}
\end{figure}

We present the derived mass spectra in Fig. \ref{mf}.
As expected, the two clusters show clear differences.
NGC 188 displays a strong deficit of low-mass stars, whereas M 67 shows an increasing number of stars toward lower masses.
Note that we excluded the most massive stars when constructing the mass spectra.
Including them would produce local maxima at 1.18 $M_\odot$ for NGC 188 and at 1.35 $M_\odot$ for M 67.

In addition, we derived mass spectra separately for the cores and halos of both clusters.
A meaningful comparison between the core and halo spectra requires probability density distributions, i.e., spectra normalized to unity.
For this purpose, we need the number of stars used to construct the mass spectrum.
We obtain this number by integrating the mass spectrum (and/or the brightness function):
\begin{equation}
N=\int\limits_{m_1}^{m_2}\psi(m)dm=\int\limits_{G_1}^{G_2}\varphi(G)dG \; ,
\label{starnum}
\end{equation}
\noindent where the stellar masses within the integral over the mass spectrum correspond to the apparent magnitudes within the integral over the brightness function.
Note that this does not represent the total number of stars in the cluster within the studied magnitude range, as the brightest stars are excluded.
To obtain the total number of stars in the cluster over the magnitude interval used to construct the brightness function, we integrate the brightness function over this interval:
\begin{equation}
N_c=\int\limits_{G_{min}}^{G_{max}}\varphi(G)dG \; ,
\label{totstarnum}
\end{equation}

We compare the normalized mass spectra of the cores and halos in Fig. \ref{mfnorm}.
We show the core spectra with pink lines and the halo spectra with blue lines.
In both clusters, low-mass stars dominate the halo regions.
For higher-mass stars, the differences in the distributions are less pronounced, although M 67 shows a relative deficit of massive stars in the halo, while NGC 188 exhibits a relative excess.
This difference is not very significant, since the most massive cluster stars were excluded from the mass spectra (see above).
Additionally, in the halo of NGC 188, compared to its core, we observe a relative deficit of stars with masses between 0.66 and 0.9 $M_\odot$.
In NGC 188, the halo and core do not differ significantly in stellar population within the mass range 0.85–1.2 $M_\odot$.

\begin{figure}[h]
\includegraphics[width=0.48\textwidth]{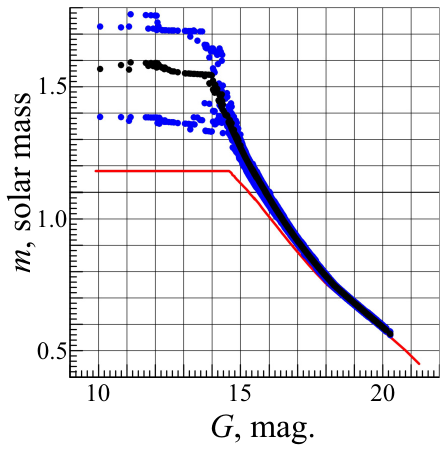}
\caption{Mass vs. apparent magnitude for NGC 188. Black points --- median mass (50th percentile), blue points --- 16th and 84th percentiles from \cite{H&R2024}, red line --- relation used in this work.
}
\label{m-G_HR}
\end{figure}

\begin{figure}[h]
\includegraphics[width=1\textwidth]{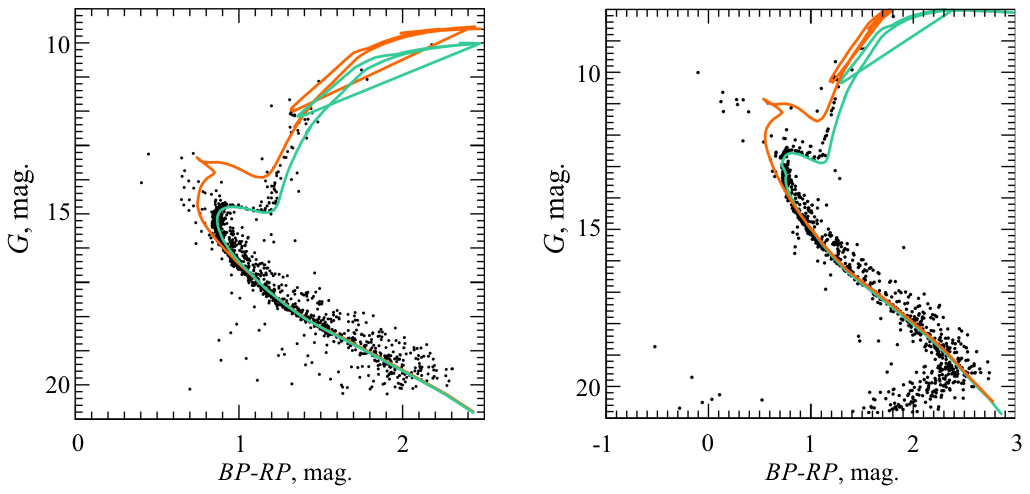}
\caption{Comparison of isochrones for cluster ages from \cite{Dias+2021} (green line) and \cite{H&R2023} (orange line). Left: NGC 188; black points --- stars from \cite{H&R2023} sample. Right: M 67; black points --- stars within 30 arcmin from the cluster center, selected using proper motion data (\ref{interval_1_67}). }
\label{izo_compar}
\end{figure}

We compared the derived mass spectra with those obtained from the samples of probable cluster members in \cite{H&R2024}.
This comparison is straightforward, as \cite{H&R2024} provides the individual stellar masses in a table.
We found that the mass spectra differ significantly, primarily in the range of stellar masses.
The mass range in \cite{H&R2024} is substantially wider.
To investigate the reason for this discrepancy, we constructed the mass–apparent magnitude relations using the same samples from \cite{H&R2024}.
For NGC 188, this relation appears in Fig. \ref{m-G_HR} as black points, representing the median mass (50th percentile), with the 16th and 84th percentiles shown as blue points.
For comparison, we show our adopted relation with a red line.
The mass–apparent magnitude relation from \cite{H&R2024} deviates strongly from ours, resembling the relation expected for a younger cluster.

Indeed, \cite{H&R2023,H&R2024} adopt logarithmic ages of $\log t = 9.42$ for NGC 188 and $\log t = 9.23$ for M 67, in contrast to the catalog values from \cite{Dias+2021}, which are $\log t = 9.79 \pm 0.03$ and $\log t = 9.58 \pm 0.03$, respectively (ages $t$ in years).
Fig. \ref{izo_compar} compares PARSEC \cite{PARSEC} isochrones constructed for the ages from \cite{H&R2023} and \cite{Dias+2021}.
The isochrone corresponding to the ages from Hunt \& Reffert \cite{H&R2023} shows a noticeable offset from the observed positions of cluster stars on the color–magnitude diagram.
In contrast, the isochrone for the Dias et al. \cite{Dias+2021} ages provides a closer agreement with the observed stellar positions.
The deviation in the red giant branch can be attributed to the fact that PARSEC isochrones do not allow variation of helium content in stellar atmospheres.
According to \cite{H&R2023,H&R2024}, the age determination errors for old clusters in their catalog may arise because the algorithm erroneously identifies blue straggler stars as turnoff stars.
Recent age estimates —-- $\log t = 9.81$ for NGC 188 \cite{Yakut+2025} and $\log t = 9.60$ for M 67 \cite{Reyes+2024} --— are very close to the values listed in the Dias et al. catalog \cite{Dias+2021}.

\section{Comparison with Hunt \& Reffert Samples, Cumulative Density Distributions}

\begin{table}
\caption{Comparison of Hunt \& Reffert \cite{H&R2023} samples with the samples obtained in the present work}
\label{tab:comparison_HR}
\begin{tabular}{c|c|c|c|c|c|c|c|c|c}
\hline
Cluster   & Astrometric &$G_{lim}$&    $R_c$  &  $N$  & $N_c$ & $P$   & $N_{HR}$ &  $N_{HR}$ & Common    \\
          & constraints &         &           &       &       &       &          & outside   & with H\&R \\
          &             &         &           &       &       &       &          &   $R_c$   &           \\
\hline
          &             & mag     &  arcmin   &       &       &       &          &           &           \\
\hline
    1     &     2       &    3    &     4     &   5   &   6   &   7   &     8    &     9     &   10   \\
\hline
NGC 188   &    (1)      &   18    &  56$\pm$4 &  1158 &   948 &  0.82 &   964    &     2     &  961   \\
          &    (1)      &   20    &  56$\pm$4 &  1846 &  1366 &  0.74 &  1324    &     3     & 1317   \\
          &    (1)      &   21    &  56$\pm$4 &  2056 &  1430 &  0.70 &  1338    &     4     & 1330   \\
          &    (2)      &   18    &  70$\pm$5 &   948 &   905 &  0.95 &   964    &     0     &  903   \\ 
          &    (2)      &   20    &  70$\pm$5 &  1243 &  1162 &  0.92 &  1324    &     0     & 1156   \\
          &    (2)      &   21    &  70$\pm$5 &  1268 &  1169 &  0.92 &  1338    &     0     & 1158   \\
\hline
M 67      &    (3)      &   18    &  59$\pm$3 &  1399 &  1174 &  0.84 &  1279    &    63     & 1216   \\
          &    (3)      &   21    &  94$\pm$8 &  3395 &  2022 &  0.60 &  1844    &     2     & 1842   \\
          &    (4)      &   21    & 133$\pm$5 &  1447 &  1400 &  0.97 &  1844    &     0     & 1340   \\
\hline
\end{tabular}
\end{table}

It is instructive to compare the samples of probable cluster members from \cite{H&R2023,H&R2024} with the samples we obtained using proper motion and parallax constraints (\ref{interval_1_188})–(\ref{interval_2_67}).
We present the results of this comparison in Table \ref{tab:comparison_HR}.
The first column lists the cluster name; the second column indicates the adopted constraints on astrometric parameters; the third column gives the limiting apparent magnitude; and the fourth column shows the cluster radius $R_c$ in arcminutes, determined from the radial surface density profile (see Section III).
The fifth column reports the number of stars located within a circle of radius $R_c$ under the conditions listed in the previous columns.
The sixth column gives the number of cluster stars within the circle of radius $R_c$, computed as the difference between the total number of stars in the circle and the number of stars in the surrounding ring with inner radius $R_c$ and outer radius $R_c \sqrt{2}$.
The seventh column presents the group membership probability $P$ for stars within the circle of radius $R_c$, determined using the uniform field method \cite{Danilov2020}.
This method assumes that field stars are distributed uniformly with a mean density equal to the stellar density in the comparison ring:
\begin{equation}
P=\frac{N_c}{N} \; .
\label{prob}
\end{equation}

\noindent In other words, $P$ represents the probability that a randomly selected star within the circle of radius $R_c$ belongs to the cluster.

The eighth column of Table \ref{tab:comparison_HR} lists the number of Hunt \& Reffert \cite{H&R2023} stars below the given limiting magnitude, while the ninth column shows the number of stars from this sample located outside the circle of radius $R_c$.
The tenth column gives the number of stars common to our sample (column 5 in Table \ref{tab:comparison_HR}) and the sample from \cite{H&R2023}.

From the data in Table \ref{tab:comparison_HR}, we can draw the following conclusions.
First, the Hunt \& Reffert \cite{H&R2023} samples are excellent.
Even stars with low membership probabilities in their sample have group probabilities, determined using the uniform field method, of at least 0.7 for NGC 188 and at least 0.6 for M 67.
Second, very simple star selection methods can yield results only slightly worse than those obtained with a more sophisticated approach (authors of \cite{H&R2023} use the HDBSCAN method applied in a multidimensional parameter space).
Moreover, in the case of NGC 188, when selecting stars using (\ref{interval_1_188}) and $G<20,21$ mag, and for M 67 using (\ref{interval_1_67}) and $G<21$ mag, our method identifies a larger number of cluster stars compared to \cite{H&R2023}.
This can be explained by the fact that the astrometric parameter constraints we applied are very broad, allowing an inclusion of some extra stars into our sample that were excluded by the methodology of \cite{H&R2023}.

It should be noted that our study and the works of \cite{H&R2023,H&R2024} use different definitions of probability.
In our work, the probability represents the chance that a randomly selected star belongs to the cluster.
In Hunt \& Reffert \cite{H&R2023,H&R2024}, the probability refers to the likelihood that a specific star is a cluster member.

\begin{table}
\caption{Intervals of stellar magnitudes and masses for comparison of cumulative density distributions}
\label{tab:mass_int}
\begin{tabular}{c|c|c|c|c|c|c|c}
\hline
Cluster & Interval & Line color & Magnitude     & Mass          & $r_{\rm max}$ & $r_{\rm min}$ & $N$ \\
        & number   & in Fig.11  & interval      & interval      & arcmin        & arcmin        &     \\
\hline
        &          &            & mag           & $M_\odot$     & arcmin        & arcmin        &     \\
\hline
 1      & 2        & 3          & 4             & 5             & 6             & 7             & 8   \\
\hline
NGC 188 & 1        & red        & $<15.4$       & $[1.1;1.2]$   & 54.84         & 0.27          & 299 \\
        & 2        & pink       & $[15.4;16.8]$ & $[0.9;1.1]$   & 69.62         & 0.30          & 383 \\
        & 3        & green      & $[16.8;18.6]$ & $[0.7;0.9]$   & 55.27         & 0.20          & 418 \\
        & 4        & blue       & $[18.6;20.3]$ & $[0.55;0.7]$  & 68.38         & 0.73          & 238 \\
\hline
M 67    & 1        & red        & $<12.5$       & $[1.3;1.37]$  & 47.46         & 0.82          &  84 \\
        & 2        & pink       & $[12.5;14.5]$ & $[1.0;1.3]$   & 95.02         & 0.12          & 437 \\
        & 3        & green      & $[14.5;16.8]$ & $[0.7;1.0]$   &100.87         & 0.27          & 513 \\
        & 4        & blue       & $[16.8;19.0]$ & $[0.5;0.7]$   & 87.64         & 0.52          & 499 \\
        & 5        & orange     & $[19.0;20.6]$ & $[0.34;0.5]$  & 86.62         & 0.69          & 311 \\
\hline
\end{tabular}
\end{table}

\begin{figure}[h]
\includegraphics[width=1\textwidth]{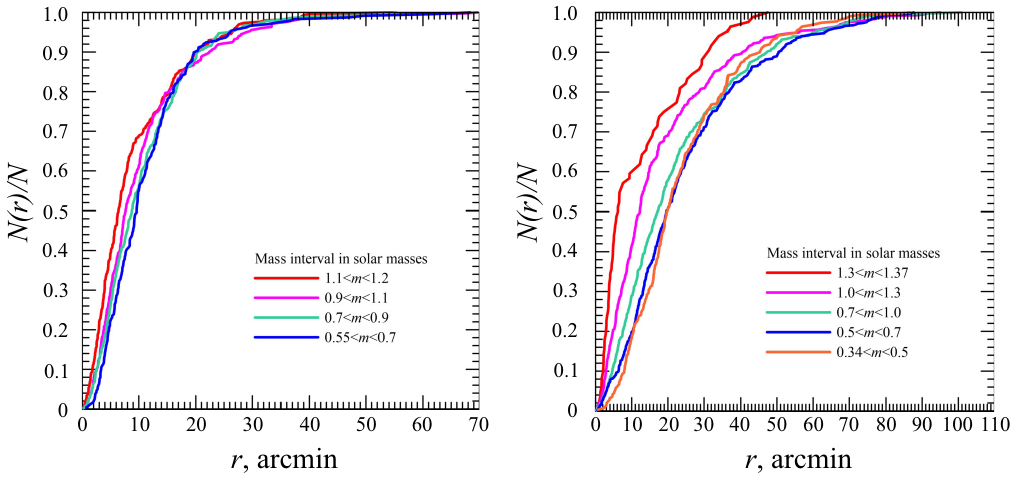}
\caption{Cumulative density distributions for stars of different masses. Left: NGC 188, right: M 67. For NGC 188, 1 arcmin corresponds to 0.54 pc; for M 67, 1 arcmin corresponds to 0.25 pc.}
\label{cumulative}
\end{figure}

Given the high quality of the \cite{H&R2023} samples, we decided to investigate the distribution of stars of different masses (or apparent magnitudes) in NGC 188 and M 67 based on these samples.
The aim of this analysis is to revisit the results of \cite{Tinsley&King1976,McClure&Twarog1977} on the distribution of red giants in NGC 188 and M 67 using modern data.
To this end, we constructed cumulative distributions of stars within different magnitude (and, correspondingly, mass) intervals.
The selected intervals are listed in Table \ref{tab:mass_int}, and the resulting distributions appear in Fig. \ref{cumulative}.
In Table \ref{tab:mass_int}, the first column gives the cluster name, the second column indicates the interval number of magnitudes and masses, the third column shows the line color in Fig. \ref{cumulative} corresponding to the interval, the fourth column gives the magnitude interval, the fifth column is the stellar mass interval in solar masses, the sixth and seventh columns report the maximum and minimum distances of stars in the interval from the cluster center in arcminutes (for NGC 188, 1 arcmin corresponds to 0.54 pc, for M 67 — 0.25 pc), and the eighth column lists the number of stars in the interval.
We determined the mass interval boundaries using the mass – apparent magnitude relations derived in the present study.

Fig. \ref{cumulative} shows that in M 67 stars above the main-sequence turnoff (red line in the right panel) concentrate toward the cluster center more strongly than any other stars.
This well-known mass segregation has been reported in numerous studies. We highlight this in the context of \cite{McClure&Twarog1977,Tinsley&King1976}, which noted a broader distribution of red giants in NGC 188 and M 67 (see the Introduction for discussion).
In NGC 188, stars above the main-sequence turnoff (red line in the left panel) also show stronger central concentration, but only up to approximately $r \approx 15$ arcminutes; beyond this radius, the distribution of the most massive stars is indistinguishable from that of lower-mass stars.
Overall, the difference in the distributions of stars of different masses is less pronounced in NGC 188 than in M 67.
Moreover, stars in the upper main sequence of NGC 188 (pink line in the left panel) appear somewhat more widely distributed beyond $r \approx 18$ arcmin compared to stars in other mass intervals.
Our results agree well with those reported by \cite{Alvarez-Baena+2024} (see Fig. 4 in that work), which also show that mass segregation in NGC 188 is much weaker than in M 67.

We compared the derived cumulative distributions using the Kolmogorov–Smirnov test.
This comparison revealed significant differences in the distributions for almost all stellar groups in the studied clusters, except for the least massive stars.
In M 67, the distributions of groups 4 and 5 do not differ significantly, with a probability value of $p \simeq 5.6 \cdot 10^{-2}$.
In NGC 188, stars in groups 2 and 3 ($p \simeq 9.0 \cdot 10^{-2}$) and groups 3 and 4 ($p \simeq 8.3 \cdot 10^{-2}$) do not differ significantly.
For all other groups, the probability values are very small, ranging from $p \approx 10^{-4}$ to values smaller by several orders of magnitude.
It should be noted that these results refer to comparisons of cumulative distributions over the entire radial interval from the cluster center.
Consequently, regions where the distributions differ most strongly carry the greatest weight in the test.

\section{Velocity Dispersion of Stars in Clusters}

Thanks to the high photometric precision in the Gaia catalogs, the color–magnitude diagrams of NGC 188 and M 67 clearly reveal the sequences of single and unresolved binary stars.
We therefore decided to compare the velocity dispersions, derived from proper motions, for single and unresolved binary stars.
When scanning the sky, the Gaia instrument measures not the center-of-mass coordinates of a binary system but the coordinates of its photocenter.
As a result, the velocity dispersions derived from proper motions differ between single and unresolved binary stars, with binaries expected to exhibit larger dispersions.
This issue has been discussed, for example, in \cite{Pang+2023}.

It is useful to clarify a few points regarding terminology.
In stellar astronomy, the velocity dispersion refers to the square root of the mean squared velocities of stars relative to the cluster centroid.
In mathematical statistics, the term dispersion (variance) denotes the second central moment of a random variable.
The mean squared velocity of stars relative to the cluster centroid corresponds, in statistical terms, to the biased estimate of the variance.
For a sufficiently large number of stars, the difference between biased and unbiased variance estimates can be neglected.
In this case, the mean squared velocity approximates the statistical variance, and the velocity dispersion in the astronomical sense approximately equals the standard deviation (the square root of the variance) in the statistical sense.

\begin{figure}[h]
\includegraphics[width=1\textwidth]{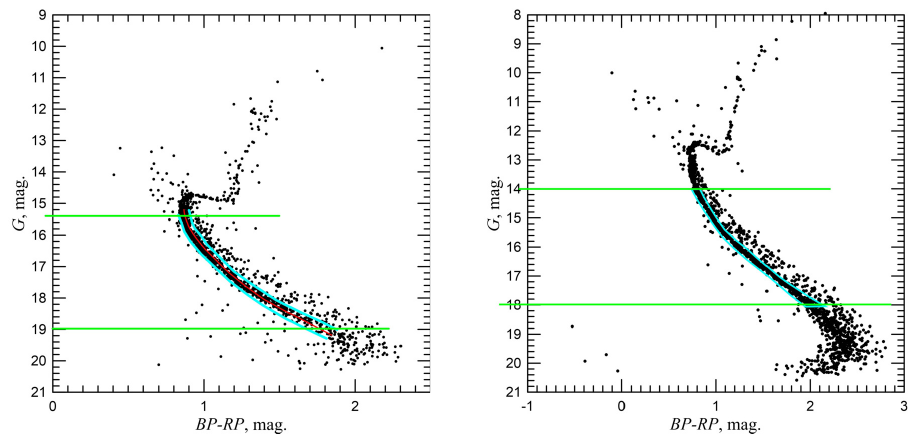}
\caption{Selection of single and unresolved binary stars for velocity dispersion determination. Green lines --- magnitude limits, blue lines --- boundaries of the single-star region. Left: NGC 188, right: M 67.}
\label{sing-bin}
\end{figure}

The selection of stars used to compute the velocity dispersion is shown in Fig. \ref{sing-bin}.
We used the samples from \cite{H&R2023}.
Green lines indicate the boundaries of the magnitude intervals.
Blue lines mark the region predominantly occupied by single stars. These lines were drawn by eye.
More accurately, the boundaries of the region of single stars should be determined using empirical isochrones for binary stars with a small mass ratio of the components \cite{Varya_VAK2024}.
However, this would not significantly affect the results of the present study, as it would reclassify only a very small number of stars between the single and binary samples.
The unresolved binaries (and higher-order multiples) with components on the main sequence lie to the right of this region.
To the left of this region, the unresolved binaries with a white dwarf component may be located \cite{WD_bin}.
Unresolved binaries with low mass ratios can also fall within the area bounded by the blue lines.
Unfortunately, separating such binaries from single stars is extremely difficult.

To compute the mean proper motions and their standard deviations, we used only stars from \cite{H&R2023} with membership probabilities greater than 0.5.
We also excluded halo stars from the calculations, as their motions may be affected by the Galactic gravitational field (the adopted cluster core radii are given in Section IV).
We did not account for proper motion errors, since the \cite{H&R2023} sample includes only stars with reliably determined parallaxes and proper motions.
We applied the standard formulas for the mean and standard deviation for values of an equal weight.

The results of the calculations are presented in Table \ref{tab:vel_disp}.
In this table, the first column lists the cluster name, the second column indicates the sample (single or binary stars), the third column gives the number of stars in the sample, the fourth column shows the mean proper motion in right ascension in milliarcseconds per year, the fifth column gives the standard deviation of the proper motion in right ascension in milliarcseconds per year, and the sixth column lists the variance of the proper motion in right ascension in squared milliarcseconds per year. Columns 7, 8, and 9 provide the same quantities for proper motion in declination as those in columns 4, 5, and 6, respectively.

\begin{table}
\caption{Mean values, standard deviations, and variances of proper motions for samples of single and binary stars}
\label{tab:vel_disp}
\begin{tabular}{c|c|c|c|c|c|c|c|c}
\hline
Cluster & Sample & Number of stars & $\overline{\mu_\alpha}$ & $\sigma_{\mu_\alpha}$ & $\sigma^2_{\mu_\alpha}$ & $\overline{\mu_\delta}$ & $\sigma_{\mu_\delta}$ & $\sigma^2_{\mu_\delta}$ \\
\hline
        &        &                 & mas/yr & mas/yr & (mas/yr)$^2$ & mas/yr & mas/yr & (mas/yr)$^2$ \\
\hline
  1     & 2      & 3               & 4       & 5       & 6             & 7       & 8       & 9 \\
\hline
NGC 188 & single & 354             & -2.31   & 0.10    & 0.0100$\pm$0.0008 & -1.01 & 0.11 & 0.0121$\pm$0.0009 \\
        & binary & 105             & -2.32   & 0.13    & 0.017$\pm$0.002  & -1.00 & 0.13 & 0.017$\pm$0.002  \\
\hline
M 67    & single & 340             & -10.94  & 0.17    & 0.029$\pm$0.002  & -2.92 & 0.17 & 0.029$\pm$0.002  \\
        & binary & 120             & -10.95  & 0.21    & 0.044$\pm$0.006  & -2.89 & 0.22 & 0.048$\pm$0.006  \\
\hline
\end{tabular}
\end{table}

Table \ref{tab:vel_disp} shows that the relative changes in the mean proper motions are very small.
Relative changes in the standard deviations of the proper motions (which we approximate as velocity dispersions) are considerably larger, ranging from 18 to 30\%.
Are these differences significant?
Cramer \cite{Cramer} provides a formula to estimate the variance of a variance:
\begin{equation}
D\sigma^2 \approx \frac{2\sigma^4}{N} ; .
\label{disp_disp}
\end{equation}
The standard deviations of the variances calculated using this formula (here, variance in the statistical sense) are listed in columns 6 and 9 of Table \ref{tab:vel_disp} (after the $\pm$ sign).
These standard deviations do not exceed 0.006.
Therefore, we can conclude that the variances $\sigma^2$ of the proper motions (and hence the standard deviations, i.e., the velocity dispersions) differ significantly.

Thus, to obtain the velocity dispersion (for example, for estimating the virial mass of a cluster), one should use the velocity dispersions (standard deviations) derived from the proper motions of single stars, excluding unresolved binary systems whenever possible.

\section{CONCLUSION}

In this study, we conducted a statistical analysis of two old open clusters, NGC 188 and M 67.
The main results can be summarized as follows.

\begin{enumerate}

\item To search for tidal tails, we constructed maps of the surface density of stars in the clusters under different constraints on stellar astrometric parameters (in other words, stars with varying group membership probabilities were used when constructing the density maps).
No tidal tails were detected. 
This is most likely explained by the fact that, given the large ages of these clusters, they have passed through the Galactic plane multiple times and experienced the repeated dynamical shocks. 
Consequently, any tidal tails had been disrupted and mixed with field stars.

\item In parallel with the density maps, we built radial profiles of the surface density to estimate the cluster sizes.  
We found that the cluster radii strongly depend on the magnitude limit and on the constraints applied to the astrometric parameters. The results are presented in Table \ref{tab:cluster_rad}.  
This effect is particularly pronounced in M 67, where the cluster radius changes by more than a factor of two when increasing the magnitude limit and the group membership probability threshold.  
The maximum radius of M 67, $R_c = 133 \pm 5$ arcminutes, significantly exceeds the total cluster radius reported in \cite{H&R2023,H&R2024}, defined as the distance from the cluster center to the most distant star in the sample of probable members.

\item For both clusters, we constructed brightness functions and mass spectra.  
NGC 188 and M 67 exhibit significant differences in the shapes of these functions.  
In M 67, the mass spectrum shows an almost continuous increase toward low-mass stars.  
In contrast, NGC 188 exhibits a deficit of low-mass stars relative to more massive ones, including a sharp drop in both the brightness function and the mass spectrum at the lowest stellar masses ($m < 0.6 \, M_\odot$).  
The reality of this drop is supported by the fact that, for field stars, the brightness function continues to rise down to the lower magnitude limit.  
The conclusion regarding the sharp drop in the brightness function and mass spectrum for stars with $m < 0.6 \, M_\odot$ in NGC 188 is valid, provided that the cluster region and the comparison field have the same level of incompleteness.
We also constructed mass spectra for the central parts of the clusters (their cores, in the terminology of P. N. Kholopov \cite{Kholopov_structure}) and for the cluster halos.  
The halos are more populous in low-mass stars, whereas for more massive stars, a significant difference is observed only in M 67 (the halo of this cluster contains fewer massive stars than the core).  
In NGC 188, the halo shows a weak (insignificant) relative excess of massive stars compared to the core.  
Additionally, in the halo of NGC 188 (compared to its core), there is a relative deficit of stars in the mass range $0.66$–$0.9 \, M_\odot$ (also insignificant).  
For stars in the mass range of $0.85$–$1.2 \, M_\odot$, the halo and core stellar population of NGC 188 do not differ significantly.  
The mass spectra obtained in this study were compared with those based on the data from \cite{H&R2024}.  
The comparison revealed substantial differences, which can be attributed to the underestimated cluster ages adopted in \cite{H&R2024}.

\item The stellar samples obtained in this study were compared with the samples from \cite{H&R2023,H&R2024}.  
It was found that the samples from \cite{H&R2023,H&R2024} are almost entirely contained within our samples (see the comparison results in Table \ref{tab:comparison_HR}).  
Furthermore, the group membership probabilities calculated in this study using the uniform field method are no less than 0.6.  
Thus, stars from the \cite{H&R2023,H&R2024} samples, even those with low membership probability ($<0.5$), have high group membership probabilities according to our calculations.  
This indicates the very high quality of the \cite{H&R2023,H&R2024} samples.  
We emphasize that the simple selection method employed in this work produced samples that include virtually all stars from \cite{H&R2023,H&R2024}.  
However, our samples also contain some field stars.  
Under certain selection criteria (see Table \ref{tab:comparison_HR}), the statistical method used here yields a larger number of cluster stars than the method applied in \cite{H&R2023,H&R2024}.

\item The samples from \cite{H&R2023,H&R2024} were used to construct cumulative stellar distributions in the clusters to reassess, with modern data, the results of \cite{McClure&Twarog1977,Tinsley&King1976}.  
These earlier studies, based on photographic observations, concluded that red giants in these clusters are more widely distributed than stars in the upper main sequence.  
The authors of \cite{McClure&Twarog1977,Tinsley&King1976} attributed this to red giants having lost significant mass and redistributed within the clusters.  
It should be noted that, according to modern data (PARSEC isochrones \cite{PARSEC}), red giants remain the most massive stars in the studied clusters.  
The cumulative distributions show that the stellar distributions in NGC 188 and M 67 differ noticeably.  
In M 67, red giants are much more centrally concentrated than other stars.  
For other mass groups in M 67, more massive stars also show a greater central concentration.  
In NGC 188, red giants are more concentrated toward the center only in the central region of the cluster (up to approximately $r\approx15$ arcminutes), beyond which their distribution is hardly distinguishable from that of other mass groups.  
Stars in the upper main sequence of NGC 188 are slightly more widely distributed than other stars in the outer region of the cluster (beyond $r\approx18$ arcminutes).  
Kolmogorov–Smirnov tests applied to the cumulative distributions indicate a significant difference in almost all stellar groups in the studied clusters (except the lowest-mass stars).  
However, this comparison is based on distributions across all distances from the cluster center, with the central regions contributing more weight.

\item This study emphasizes the terminological differences between stellar astronomy and mathematical statistics.  
In stellar astronomy, the term `velocity dispersion' approximately corresponds to the standard deviation of velocities in mathematical statistics.  
Conversely, the concept of dispersion (`variance') in statistics can be roughly associated with the mean square of stellar velocities relative to the cluster centroid.  
This discrepancy in terminology should be taken into account.

\item In this study, we obtained estimates of the standard deviations of stellar proper motions relative to their mean values (these quantities are used to derive the velocity dispersion in km/s).  
It was found that the dispersions and standard deviations of proper motions for samples of single and unresolved binary stars differ significantly.  
Standard deviations of proper motions for unresolved binaries are significantly larger than those for single stars, with relative differences ranging from 18\% to 30\%.  
Therefore, when determining the mean square velocity and velocity dispersion in a cluster (in the stellar-astronomical sense), it is advisable, whenever possible, to use only single stars.

\end{enumerate}

What causes such a pronounced difference between the clusters NGC 188 and M 67?
Most likely, the reason lies in their different dynamical histories.
NGC 188 has a higher $z$–coordinate than M 67.
Consequently, it is possible that NGC 188 passed through the Galactic plane at a higher velocity and experienced a stronger dynamical shock.
As a result, the loss of low-mass stars in NGC 188 may have been more intense compared to M 67.
NGC 188 may also exhibit a higher degree of non-stationarity \cite{Danilov_Putkov_2012ARep} than M~67.
To justify conclusions about the cluster’s non-stationarity, a detailed analysis of the velocity field is required (see, e.g., \cite{Danilov2020}).
Unfortunately, for NGC 188 and M 67, the tangential velocity errors (even using Gaia data) are too large for a detailed study due to their substantial distances from the Sun.
It is possible that the final Gaia mission data release will enable such an investigation, at least for M 67.

\section*{Funding}
This work was supported by the Ministry of Science and Higher Education of the Russian Federation, project FEUZ-2023-0019.

\begin{acknowledgments}
This study used data from the European Space Agency (ESA) Gaia mission (https://www.cosmos.esa.int/gaia) processed by the Gaia Data Processing and Analysis Consortium (DPAC, https://www.cosmos.esa.int/web/gaia/dpac/consortium). Funding for DPAC has been provided by national institutions, particularly those participating in the Gaia Multilateral Agreement.
\end{acknowledgments}

\section*{Conflict of Interest}
The authors declare no conflict of interest.

\bibliographystyle{maik}
\bibliography{Seleznev_188_67}

\begin{thebibliography}{52}
\expandafter\ifx\csname natexlab\endcsname\relax\def\natexlab#1{#1}\fi
\expandafter\ifx\csname bibnamefont\endcsname\relax
  \def\bibnamefont#1{#1}\fi
\expandafter\ifx\csname bibfnamefont\endcsname\relax
  \def\bibfnamefont#1{#1}\fi
\expandafter\ifx\csname citenamefont\endcsname\relax
  \def\citenamefont#1{#1}\fi
\expandafter\ifx\csname url\endcsname\relax
  \def\url#1{\texttt{#1}}\fi
\expandafter\ifx\csname urlprefix\endcsname\relax\def\urlprefix{URL }\fi
\providecommand{\bibinfo}[2]{#2}
\providecommand{\eprint}[2][]{\url{#2}}

\bibitem[{\citenamefont{{Danilov} and {Seleznev}}(2020)}]{Danilov2020}
\bibinfo{author}{\bibfnamefont{V.~M.} \bibnamefont{{Danilov}}}
  \bibnamefont{and} \bibinfo{author}{\bibfnamefont{A.~F.}
  \bibnamefont{{Seleznev}}}, \bibinfo{journal}{Astrophysical Bulletin}
  \textbf{\bibinfo{volume}{75}}, \bibinfo{pages}{407} (\bibinfo{year}{2020}).

\bibitem[{\citenamefont{{Gaia Collaboration}
  \emph{et~al.}}(2016)\citenamefont{{Gaia Collaboration}, {Prusti}, {de
  Bruijne}, {Brown}, {Vallenari}, {Babusiaux}, {Bailer-Jones}, {Bastian},
  {Biermann}, {Evans} \emph{et~al.}}}]{Gaia}
\bibinfo{author}{\bibnamefont{{Gaia Collaboration}}},
  \bibinfo{author}{\bibfnamefont{T.}~\bibnamefont{{Prusti}}},
  \bibinfo{author}{\bibfnamefont{J.~H.~J.} \bibnamefont{{de Bruijne}}},
  \bibinfo{author}{\bibfnamefont{A.~G.~A.} \bibnamefont{{Brown}}},
  \bibinfo{author}{\bibfnamefont{A.}~\bibnamefont{{Vallenari}}},
  \bibinfo{author}{\bibfnamefont{C.}~\bibnamefont{{Babusiaux}}},
  \bibinfo{author}{\bibfnamefont{C.~A.~L.} \bibnamefont{{Bailer-Jones}}},
  \bibinfo{author}{\bibfnamefont{U.}~\bibnamefont{{Bastian}}},
  \bibinfo{author}{\bibfnamefont{M.}~\bibnamefont{{Biermann}}},
  \bibinfo{author}{\bibfnamefont{D.~W.} \bibnamefont{{Evans}}},
  \bibnamefont{\emph{et~al.}}, \bibinfo{journal}{Astronomy and Astrophysics}
  \textbf{\bibinfo{volume}{595}}, \bibinfo{eid}{A1} (\bibinfo{year}{2016}).

\bibitem[{\citenamefont{{Gaia Collaboration}
  \emph{et~al.}}(2018)\citenamefont{{Gaia Collaboration}, {Brown}, {Vallenari},
  {Prusti}, {de Bruijne}, {Babusiaux}, {Bailer-Jones}, {Biermann}, {Evans},
  {Eyer} \emph{et~al.}}}]{GaiaDR2}
\bibinfo{author}{\bibnamefont{{Gaia Collaboration}}},
  \bibinfo{author}{\bibfnamefont{A.~G.~A.} \bibnamefont{{Brown}}},
  \bibinfo{author}{\bibfnamefont{A.}~\bibnamefont{{Vallenari}}},
  \bibinfo{author}{\bibfnamefont{T.}~\bibnamefont{{Prusti}}},
  \bibinfo{author}{\bibfnamefont{J.~H.~J.} \bibnamefont{{de Bruijne}}},
  \bibinfo{author}{\bibfnamefont{C.}~\bibnamefont{{Babusiaux}}},
  \bibinfo{author}{\bibfnamefont{C.~A.~L.} \bibnamefont{{Bailer-Jones}}},
  \bibinfo{author}{\bibfnamefont{M.}~\bibnamefont{{Biermann}}},
  \bibinfo{author}{\bibfnamefont{D.~W.} \bibnamefont{{Evans}}},
  \bibinfo{author}{\bibfnamefont{L.}~\bibnamefont{{Eyer}}},
  \bibnamefont{\emph{et~al.}}, \bibinfo{journal}{Astronomy and Astrophysics}
  \textbf{\bibinfo{volume}{616}}, \bibinfo{eid}{A1} (\bibinfo{year}{2018}).

\bibitem[{\citenamefont{{Gaia Collaboration}
  \emph{et~al.}}(2023)\citenamefont{{Gaia Collaboration}, {Vallenari}, {Brown},
  {Prusti}, {de Bruijne}, {Arenou}, {Babusiaux}, {Biermann}, {Creevey},
  {Ducourant} \emph{et~al.}}}]{GaiaDR3}
\bibinfo{author}{\bibnamefont{{Gaia Collaboration}}},
  \bibinfo{author}{\bibfnamefont{A.}~\bibnamefont{{Vallenari}}},
  \bibinfo{author}{\bibfnamefont{A.~G.~A.} \bibnamefont{{Brown}}},
  \bibinfo{author}{\bibfnamefont{T.}~\bibnamefont{{Prusti}}},
  \bibinfo{author}{\bibfnamefont{J.~H.~J.} \bibnamefont{{de Bruijne}}},
  \bibinfo{author}{\bibfnamefont{F.}~\bibnamefont{{Arenou}}},
  \bibinfo{author}{\bibfnamefont{C.}~\bibnamefont{{Babusiaux}}},
  \bibinfo{author}{\bibfnamefont{M.}~\bibnamefont{{Biermann}}},
  \bibinfo{author}{\bibfnamefont{O.~L.} \bibnamefont{{Creevey}}},
  \bibinfo{author}{\bibfnamefont{C.}~\bibnamefont{{Ducourant}}},
  \bibnamefont{\emph{et~al.}}, \bibinfo{journal}{Astronomy and Astrophysics}
  \textbf{\bibinfo{volume}{674}}, \bibinfo{eid}{A1} (\bibinfo{year}{2023}).

\bibitem[{\citenamefont{{Cantat-Gaudin} and
  {Casamiquela}}(2024)}]{Cantat-Gaudin2024}
\bibinfo{author}{\bibfnamefont{T.}~\bibnamefont{{Cantat-Gaudin}}}
  \bibnamefont{and}
  \bibinfo{author}{\bibfnamefont{L.}~\bibnamefont{{Casamiquela}}},
  \bibinfo{journal}{New Astronomy Reviews} \textbf{\bibinfo{volume}{99}},
  \bibinfo{eid}{101696} (\bibinfo{year}{2024}).

\bibitem[{\citenamefont{{Hunt} and {Reffert}}(2023)}]{H&R2023}
\bibinfo{author}{\bibfnamefont{E.~L.} \bibnamefont{{Hunt}}} \bibnamefont{and}
  \bibinfo{author}{\bibfnamefont{S.}~\bibnamefont{{Reffert}}},
  \bibinfo{journal}{Astronomy and Astrophysics} \textbf{\bibinfo{volume}{673}},
  \bibinfo{eid}{A114} (\bibinfo{year}{2023}).

\bibitem[{\citenamefont{{Hunt} and {Reffert}}(2024)}]{H&R2024}
\bibinfo{author}{\bibfnamefont{E.~L.} \bibnamefont{{Hunt}}} \bibnamefont{and}
  \bibinfo{author}{\bibfnamefont{S.}~\bibnamefont{{Reffert}}},
  \bibinfo{journal}{Astronomy and Astrophysics} \textbf{\bibinfo{volume}{686}},
  \bibinfo{eid}{A42} (\bibinfo{year}{2024}).

\bibitem[{\citenamefont{{Danilov}}(2021{\natexlab{a}})}]{Danilov2021a}
\bibinfo{author}{\bibfnamefont{V.~M.} \bibnamefont{{Danilov}}},
  \bibinfo{journal}{Astrophysical Bulletin} \textbf{\bibinfo{volume}{76}},
  \bibinfo{pages}{55} (\bibinfo{year}{2021}{\natexlab{a}}).

\bibitem[{\citenamefont{{Danilov}}(2021{\natexlab{b}})}]{Danilov2021b}
\bibinfo{author}{\bibfnamefont{V.~M.} \bibnamefont{{Danilov}}},
  \bibinfo{journal}{Astrophysical Bulletin} \textbf{\bibinfo{volume}{76}},
  \bibinfo{pages}{269} (\bibinfo{year}{2021}{\natexlab{b}}).

\bibitem[{\citenamefont{{Tagaev} and
  {Seleznev}}(2025{\natexlab{a}})}]{Tagaev3532}
\bibinfo{author}{\bibfnamefont{D.~I.} \bibnamefont{{Tagaev}}} \bibnamefont{and}
  \bibinfo{author}{\bibfnamefont{A.~F.} \bibnamefont{{Seleznev}}},
  \bibinfo{journal}{Astronomy Reports} \textbf{\bibinfo{volume}{69}},
  \bibinfo{pages}{457} (\bibinfo{year}{2025}{\natexlab{a}}).

\bibitem[{\citenamefont{{Tagaev} and
  {Seleznev}}(2025{\natexlab{b}})}]{Tagaev_stat}
\bibinfo{author}{\bibfnamefont{D.~I.} \bibnamefont{{Tagaev}}} \bibnamefont{and}
  \bibinfo{author}{\bibfnamefont{A.~F.} \bibnamefont{{Seleznev}}},
  \bibinfo{journal}{Astronomy Reports, accepted}
  (\bibinfo{year}{2025}{\natexlab{b}}), \eprint{2509.10435}.

\bibitem[{\citenamefont{{Tinsley} and {King}}(1976)}]{Tinsley&King1976}
\bibinfo{author}{\bibfnamefont{B.~M.} \bibnamefont{{Tinsley}}}
  \bibnamefont{and} \bibinfo{author}{\bibfnamefont{I.~R.}
  \bibnamefont{{King}}}, \bibinfo{journal}{Astronomical Journal}
  \textbf{\bibinfo{volume}{81}}, \bibinfo{pages}{835} (\bibinfo{year}{1976}).

\bibitem[{\citenamefont{{McClure} and {Twarog}}(1977)}]{McClure&Twarog1977}
\bibinfo{author}{\bibfnamefont{R.~D.} \bibnamefont{{McClure}}}
  \bibnamefont{and} \bibinfo{author}{\bibfnamefont{B.~A.}
  \bibnamefont{{Twarog}}}, \bibinfo{journal}{Astrophysical Journal}
  \textbf{\bibinfo{volume}{214}}, \bibinfo{pages}{111} (\bibinfo{year}{1977}).

\bibitem[{\citenamefont{{Hawarden}}(1975)}]{Hawarden1975}
\bibinfo{author}{\bibfnamefont{T.~G.} \bibnamefont{{Hawarden}}},
  \bibinfo{journal}{Monthly Notices of the Royal Astronomical Society}
  \textbf{\bibinfo{volume}{173}}, \bibinfo{pages}{223} (\bibinfo{year}{1975}).

\bibitem[{\citenamefont{{Bressan} \emph{et~al.}}(2012)\citenamefont{{Bressan},
  {Marigo}, {Girardi}, {Salasnich}, {Dal Cero}, {Rubele}, and
  {Nanni}}}]{PARSEC}
\bibinfo{author}{\bibfnamefont{A.}~\bibnamefont{{Bressan}}},
  \bibinfo{author}{\bibfnamefont{P.}~\bibnamefont{{Marigo}}},
  \bibinfo{author}{\bibfnamefont{L.}~\bibnamefont{{Girardi}}},
  \bibinfo{author}{\bibfnamefont{B.}~\bibnamefont{{Salasnich}}},
  \bibinfo{author}{\bibfnamefont{C.}~\bibnamefont{{Dal Cero}}},
  \bibinfo{author}{\bibfnamefont{S.}~\bibnamefont{{Rubele}}}, \bibnamefont{and}
  \bibinfo{author}{\bibfnamefont{A.}~\bibnamefont{{Nanni}}},
  \bibinfo{journal}{Monthly Notices of the Royal Astronomical Society}
  \textbf{\bibinfo{volume}{427}}, \bibinfo{pages}{127} (\bibinfo{year}{2012}).

\bibitem[{\citenamefont{{Dias} \emph{et~al.}}(2021)\citenamefont{{Dias},
  {Monteiro}, {Moitinho}, {L{\'e}pine}, {Carraro}, {Paunzen}, {Alessi}, and
  {Villela}}}]{Dias+2021}
\bibinfo{author}{\bibfnamefont{W.~S.} \bibnamefont{{Dias}}},
  \bibinfo{author}{\bibfnamefont{H.}~\bibnamefont{{Monteiro}}},
  \bibinfo{author}{\bibfnamefont{A.}~\bibnamefont{{Moitinho}}},
  \bibinfo{author}{\bibfnamefont{J.~R.~D.} \bibnamefont{{L{\'e}pine}}},
  \bibinfo{author}{\bibfnamefont{G.}~\bibnamefont{{Carraro}}},
  \bibinfo{author}{\bibfnamefont{E.}~\bibnamefont{{Paunzen}}},
  \bibinfo{author}{\bibfnamefont{B.}~\bibnamefont{{Alessi}}}, \bibnamefont{and}
  \bibinfo{author}{\bibfnamefont{L.}~\bibnamefont{{Villela}}},
  \bibinfo{journal}{Monthly Notices of the Royal Astronomical Society}
  \textbf{\bibinfo{volume}{504}}, \bibinfo{pages}{356} (\bibinfo{year}{2021}).

\bibitem[{\citenamefont{{Mathieu} and {Geller}}(2015)}]{188_Geller_Mathieu}
\bibinfo{author}{\bibfnamefont{R.~D.} \bibnamefont{{Mathieu}}}
  \bibnamefont{and} \bibinfo{author}{\bibfnamefont{A.~M.}
  \bibnamefont{{Geller}}}, in \emph{\bibinfo{booktitle}{Astrophysics and Space
  Science Library}}, edited by \bibinfo{editor}{\bibfnamefont{H.~M.~J.}
  \bibnamefont{{Boffin}}},
  \bibinfo{editor}{\bibfnamefont{G.}~\bibnamefont{{Carraro}}},
  \bibnamefont{and} \bibinfo{editor}{\bibfnamefont{G.}~\bibnamefont{{Beccari}}}
  (\bibinfo{year}{2015}), vol. \bibinfo{volume}{413} of
  \emph{\bibinfo{series}{Astrophysics and Space Science Library}},
  p.~\bibinfo{pages}{29}.

\bibitem[{\citenamefont{{Caputo} \emph{et~al.}}(1990)\citenamefont{{Caputo},
  {Chieffi}, {Castellani}, {Collados}, {Martinez Roger}, and
  {Paez}}}]{Caputo+1990}
\bibinfo{author}{\bibfnamefont{F.}~\bibnamefont{{Caputo}}},
  \bibinfo{author}{\bibfnamefont{A.}~\bibnamefont{{Chieffi}}},
  \bibinfo{author}{\bibfnamefont{V.}~\bibnamefont{{Castellani}}},
  \bibinfo{author}{\bibfnamefont{M.}~\bibnamefont{{Collados}}},
  \bibinfo{author}{\bibfnamefont{C.}~\bibnamefont{{Martinez Roger}}},
  \bibnamefont{and} \bibinfo{author}{\bibfnamefont{E.}~\bibnamefont{{Paez}}},
  \bibinfo{journal}{Astronomical Journal} \textbf{\bibinfo{volume}{99}},
  \bibinfo{pages}{261} (\bibinfo{year}{1990}).

\bibitem[{\citenamefont{{von Hippel} and
  {Sarajedini}}(1998)}]{vonHippel_Sarajedini_1998AJ}
\bibinfo{author}{\bibfnamefont{T.}~\bibnamefont{{von Hippel}}}
  \bibnamefont{and}
  \bibinfo{author}{\bibfnamefont{A.}~\bibnamefont{{Sarajedini}}},
  \bibinfo{journal}{Astronomical Journal} \textbf{\bibinfo{volume}{116}},
  \bibinfo{pages}{1789} (\bibinfo{year}{1998}).

\bibitem[{\citenamefont{{Sarajedini}
  \emph{et~al.}}(1999)\citenamefont{{Sarajedini}, {von Hippel},
  {Kozhurina-Platais}, and {Demarque}}}]{Sarajedini+1999AJ}
\bibinfo{author}{\bibfnamefont{A.}~\bibnamefont{{Sarajedini}}},
  \bibinfo{author}{\bibfnamefont{T.}~\bibnamefont{{von Hippel}}},
  \bibinfo{author}{\bibfnamefont{V.}~\bibnamefont{{Kozhurina-Platais}}},
  \bibnamefont{and}
  \bibinfo{author}{\bibfnamefont{P.}~\bibnamefont{{Demarque}}},
  \bibinfo{journal}{Astronomical Journal} \textbf{\bibinfo{volume}{118}},
  \bibinfo{pages}{2894} (\bibinfo{year}{1999}).

\bibitem[{\citenamefont{{Bonatto} \emph{et~al.}}(2005)\citenamefont{{Bonatto},
  {Bica}, and {Santos}}}]{Bonatto+2005A&A}
\bibinfo{author}{\bibfnamefont{C.}~\bibnamefont{{Bonatto}}},
  \bibinfo{author}{\bibfnamefont{E.}~\bibnamefont{{Bica}}}, \bibnamefont{and}
  \bibinfo{author}{\bibfnamefont{J.~F.~C.} \bibnamefont{{Santos}},
  \bibfnamefont{Jr.}}, \bibinfo{journal}{Astronomy and Astrophysics}
  \textbf{\bibinfo{volume}{433}}, \bibinfo{pages}{917} (\bibinfo{year}{2005}).

\bibitem[{\citenamefont{{Elsanhoury}
  \emph{et~al.}}(2016)\citenamefont{{Elsanhoury}, {Haroon}, {Chupina},
  {Vereshchagin}, {Sariya}, {Yadav}, and {Jiang}}}]{Elsanhoury+2016NewA}
\bibinfo{author}{\bibfnamefont{W.~H.} \bibnamefont{{Elsanhoury}}},
  \bibinfo{author}{\bibfnamefont{A.~A.} \bibnamefont{{Haroon}}},
  \bibinfo{author}{\bibfnamefont{N.~V.} \bibnamefont{{Chupina}}},
  \bibinfo{author}{\bibfnamefont{S.~V.} \bibnamefont{{Vereshchagin}}},
  \bibinfo{author}{\bibfnamefont{D.~P.} \bibnamefont{{Sariya}}},
  \bibinfo{author}{\bibfnamefont{R.~K.~S.} \bibnamefont{{Yadav}}},
  \bibnamefont{and} \bibinfo{author}{\bibfnamefont{I.-G.}
  \bibnamefont{{Jiang}}}, \bibinfo{journal}{New Astronomy}
  \textbf{\bibinfo{volume}{49}}, \bibinfo{pages}{32} (\bibinfo{year}{2016}).

\bibitem[{\citenamefont{{van den Bergh}}(1957)}]{van_den_Bergh1957AJ}
\bibinfo{author}{\bibfnamefont{S.}~\bibnamefont{{van den Bergh}}},
  \bibinfo{journal}{Astronomical Journal} \textbf{\bibinfo{volume}{62}},
  \bibinfo{pages}{100} (\bibinfo{year}{1957}).

\bibitem[{\citenamefont{{Kholopov} and
  {Artyukhina}}(1965)}]{Kholopov_Artyukhina_1965SvA}
\bibinfo{author}{\bibfnamefont{P.~N.} \bibnamefont{{Kholopov}}}
  \bibnamefont{and} \bibinfo{author}{\bibfnamefont{N.~M.}
  \bibnamefont{{Artyukhina}}}, \bibinfo{journal}{Soviet Astronomy}
  \textbf{\bibinfo{volume}{8}}, \bibinfo{pages}{775} (\bibinfo{year}{1965}).

\bibitem[{\citenamefont{{Bonatto} and {Bica}}(2003)}]{Bonatto_Bica_2003A&A}
\bibinfo{author}{\bibfnamefont{C.}~\bibnamefont{{Bonatto}}} \bibnamefont{and}
  \bibinfo{author}{\bibfnamefont{E.}~\bibnamefont{{Bica}}},
  \bibinfo{journal}{Astronomy and Astrophysics} \textbf{\bibinfo{volume}{405}},
  \bibinfo{pages}{525} (\bibinfo{year}{2003}).

\bibitem[{\citenamefont{{Davenport} and
  {Sandquist}}(2010)}]{Davenport_Sandquist_2010ApJ}
\bibinfo{author}{\bibfnamefont{J.~R.~A.} \bibnamefont{{Davenport}}}
  \bibnamefont{and} \bibinfo{author}{\bibfnamefont{E.~L.}
  \bibnamefont{{Sandquist}}}, \bibinfo{journal}{Astrophysical Journal}
  \textbf{\bibinfo{volume}{711}}, \bibinfo{pages}{559} (\bibinfo{year}{2010}).

\bibitem[{\citenamefont{{Chumak} \emph{et~al.}}(2010)\citenamefont{{Chumak},
  {Platais}, {McLaughlin}, {Rastorguev}, and {Chumak}}}]{Chumak+2010MNRAS}
\bibinfo{author}{\bibfnamefont{Y.~O.} \bibnamefont{{Chumak}}},
  \bibinfo{author}{\bibfnamefont{I.}~\bibnamefont{{Platais}}},
  \bibinfo{author}{\bibfnamefont{D.~E.} \bibnamefont{{McLaughlin}}},
  \bibinfo{author}{\bibfnamefont{A.~S.} \bibnamefont{{Rastorguev}}},
  \bibnamefont{and} \bibinfo{author}{\bibfnamefont{O.~V.}
  \bibnamefont{{Chumak}}}, \bibinfo{journal}{Monthly Notices of the Royal
  Astronomical Society} \textbf{\bibinfo{volume}{402}}, \bibinfo{pages}{1841}
  (\bibinfo{year}{2010}).

\bibitem[{\citenamefont{{Hurley} \emph{et~al.}}(2005)\citenamefont{{Hurley},
  {Pols}, {Aarseth}, and {Tout}}}]{Hurley+2005MNRAS}
\bibinfo{author}{\bibfnamefont{J.~R.} \bibnamefont{{Hurley}}},
  \bibinfo{author}{\bibfnamefont{O.~R.} \bibnamefont{{Pols}}},
  \bibinfo{author}{\bibfnamefont{S.~J.} \bibnamefont{{Aarseth}}},
  \bibnamefont{and} \bibinfo{author}{\bibfnamefont{C.~A.}
  \bibnamefont{{Tout}}}, \bibinfo{journal}{Monthly Notices of the Royal
  Astronomical Society} \textbf{\bibinfo{volume}{363}}, \bibinfo{pages}{293}
  (\bibinfo{year}{2005}).

\bibitem[{\citenamefont{{Silverman}}(1986)}]{Silverman1986}
\bibinfo{author}{\bibfnamefont{B.~W.} \bibnamefont{{Silverman}}},
  \emph{\bibinfo{title}{{Density estimation for statistics and data analysis}}}
  (\bibinfo{publisher}{Chapman \& Hall, London}, \bibinfo{year}{1986}).

\bibitem[{\citenamefont{{Nikiforova}
  \emph{et~al.}}(2020)\citenamefont{{Nikiforova}, {Kulesh}, {Seleznev}, and
  {Carraro}}}]{Nikiforova2020}
\bibinfo{author}{\bibfnamefont{V.~V.} \bibnamefont{{Nikiforova}}},
  \bibinfo{author}{\bibfnamefont{M.~V.} \bibnamefont{{Kulesh}}},
  \bibinfo{author}{\bibfnamefont{A.~F.} \bibnamefont{{Seleznev}}},
  \bibnamefont{and}
  \bibinfo{author}{\bibfnamefont{G.}~\bibnamefont{{Carraro}}},
  \bibinfo{journal}{Astronomical Journal} \textbf{\bibinfo{volume}{160}},
  \bibinfo{eid}{142} (\bibinfo{year}{2020}).

\bibitem[{\citenamefont{{Yeh} \emph{et~al.}}(2019)\citenamefont{{Yeh},
  {Carraro}, {Montalto}, and {Seleznev}}}]{Yeh2019}
\bibinfo{author}{\bibfnamefont{F.~C.} \bibnamefont{{Yeh}}},
  \bibinfo{author}{\bibfnamefont{G.}~\bibnamefont{{Carraro}}},
  \bibinfo{author}{\bibfnamefont{M.}~\bibnamefont{{Montalto}}},
  \bibnamefont{and} \bibinfo{author}{\bibfnamefont{A.~F.}
  \bibnamefont{{Seleznev}}}, \bibinfo{journal}{Astronomical Journal}
  \textbf{\bibinfo{volume}{157}}, \bibinfo{eid}{115} (\bibinfo{year}{2019}).

\bibitem[{\citenamefont{{Gao}}(2020)}]{Gao2020}
\bibinfo{author}{\bibfnamefont{X.}~\bibnamefont{{Gao}}},
  \bibinfo{journal}{Publications of the Astronomical Society of Japan}
  \textbf{\bibinfo{volume}{72}}, \bibinfo{eid}{47} (\bibinfo{year}{2020}).

\bibitem[{\citenamefont{{Kos}}(2024)}]{Kos2024}
\bibinfo{author}{\bibfnamefont{J.}~\bibnamefont{{Kos}}},
  \bibinfo{journal}{Astronomy and Astrophysics} \textbf{\bibinfo{volume}{691}},
  \bibinfo{eid}{A28} (\bibinfo{year}{2024}).

\bibitem[{\citenamefont{{Danilov} and {Seleznev}}(1995)}]{Danilov&Seleznev1995}
\bibinfo{author}{\bibfnamefont{V.~M.} \bibnamefont{{Danilov}}}
  \bibnamefont{and} \bibinfo{author}{\bibfnamefont{A.~F.}
  \bibnamefont{{Seleznev}}}, \bibinfo{journal}{Astronomy Reports}
  \textbf{\bibinfo{volume}{39}}, \bibinfo{pages}{295} (\bibinfo{year}{1995}).

\bibitem[{\citenamefont{{Alvarez-Baena}
  \emph{et~al.}}(2024)\citenamefont{{Alvarez-Baena}, {Carrera}, {Thompson},
  {Balaguer-Nu{\~n}ez}, {Bragaglia}, {Jordi}, {Silva-Villa}, and
  {Vallenari}}}]{Alvarez-Baena+2024}
\bibinfo{author}{\bibfnamefont{N.}~\bibnamefont{{Alvarez-Baena}}},
  \bibinfo{author}{\bibfnamefont{R.}~\bibnamefont{{Carrera}}},
  \bibinfo{author}{\bibfnamefont{H.}~\bibnamefont{{Thompson}}},
  \bibinfo{author}{\bibfnamefont{L.}~\bibnamefont{{Balaguer-Nu{\~n}ez}}},
  \bibinfo{author}{\bibfnamefont{A.}~\bibnamefont{{Bragaglia}}},
  \bibinfo{author}{\bibfnamefont{C.}~\bibnamefont{{Jordi}}},
  \bibinfo{author}{\bibfnamefont{E.}~\bibnamefont{{Silva-Villa}}},
  \bibnamefont{and}
  \bibinfo{author}{\bibfnamefont{A.}~\bibnamefont{{Vallenari}}},
  \bibinfo{journal}{Astronomy and Astrophysics} \textbf{\bibinfo{volume}{687}},
  \bibinfo{eid}{A101} (\bibinfo{year}{2024}).

\bibitem[{\citenamefont{{Seleznev}}(2016{\natexlab{a}})}]{Seleznev2016}
\bibinfo{author}{\bibfnamefont{A.~F.} \bibnamefont{{Seleznev}}},
  \bibinfo{journal}{Monthly Notices of the Royal Astronomical Society}
  \textbf{\bibinfo{volume}{456}}, \bibinfo{pages}{3757}
  (\bibinfo{year}{2016}{\natexlab{a}}).

\bibitem[{\citenamefont{{Seleznev}}(2016{\natexlab{b}})}]{KDE_OSC}
\bibinfo{author}{\bibfnamefont{A.~F.} \bibnamefont{{Seleznev}}},
  \bibinfo{journal}{Baltic Astronomy} \textbf{\bibinfo{volume}{25}},
  \bibinfo{pages}{267} (\bibinfo{year}{2016}{\natexlab{b}}).

\bibitem[{\citenamefont{{Seleznev}}(1998)}]{Seleznev_LF}
\bibinfo{author}{\bibfnamefont{A.~F.} \bibnamefont{{Seleznev}}},
  \bibinfo{journal}{Astronomy Reports} \textbf{\bibinfo{volume}{42}},
  \bibinfo{pages}{153} (\bibinfo{year}{1998}).

\bibitem[{\citenamefont{{Fabricius}
  \emph{et~al.}}(2021)\citenamefont{{Fabricius}, {Luri}, {Arenou}, {Babusiaux},
  {Helmi}, {Muraveva}, {Reyl{\'e}}, {Spoto}, {Vallenari}, {Antoja}
  \emph{et~al.}}}]{Fabricius+2021}
\bibinfo{author}{\bibfnamefont{C.}~\bibnamefont{{Fabricius}}},
  \bibinfo{author}{\bibfnamefont{X.}~\bibnamefont{{Luri}}},
  \bibinfo{author}{\bibfnamefont{F.}~\bibnamefont{{Arenou}}},
  \bibinfo{author}{\bibfnamefont{C.}~\bibnamefont{{Babusiaux}}},
  \bibinfo{author}{\bibfnamefont{A.}~\bibnamefont{{Helmi}}},
  \bibinfo{author}{\bibfnamefont{T.}~\bibnamefont{{Muraveva}}},
  \bibinfo{author}{\bibfnamefont{C.}~\bibnamefont{{Reyl{\'e}}}},
  \bibinfo{author}{\bibfnamefont{F.}~\bibnamefont{{Spoto}}},
  \bibinfo{author}{\bibfnamefont{A.}~\bibnamefont{{Vallenari}}},
  \bibinfo{author}{\bibfnamefont{T.}~\bibnamefont{{Antoja}}},
  \bibnamefont{\emph{et~al.}}, \bibinfo{journal}{Astronomy and Astrophysics}
  \textbf{\bibinfo{volume}{649}}, \bibinfo{eid}{A5} (\bibinfo{year}{2021}).

\bibitem[{\citenamefont{{Kholopov}}(1969)}]{Kholopov_structure}
\bibinfo{author}{\bibfnamefont{P.~N.} \bibnamefont{{Kholopov}}},
  \bibinfo{journal}{Soviet Astronomy} \textbf{\bibinfo{volume}{12}},
  \bibinfo{pages}{625} (\bibinfo{year}{1969}).

\bibitem[{\citenamefont{{Seleznev}
  \emph{et~al.}}(2017)\citenamefont{{Seleznev}, {Carraro}, {Capuzzo-Dolcetta},
  {Monaco}, and {Baume}}}]{4337}
\bibinfo{author}{\bibfnamefont{A.~F.} \bibnamefont{{Seleznev}}},
  \bibinfo{author}{\bibfnamefont{G.}~\bibnamefont{{Carraro}}},
  \bibinfo{author}{\bibfnamefont{R.}~\bibnamefont{{Capuzzo-Dolcetta}}},
  \bibinfo{author}{\bibfnamefont{L.}~\bibnamefont{{Monaco}}}, \bibnamefont{and}
  \bibinfo{author}{\bibfnamefont{G.}~\bibnamefont{{Baume}}},
  \bibinfo{journal}{Monthly Notices of the Royal Astronomical Siciety}
  \textbf{\bibinfo{volume}{467}}, \bibinfo{pages}{2517} (\bibinfo{year}{2017}).

\bibitem[{\citenamefont{{Bertelli}
  \emph{et~al.}}(1994)\citenamefont{{Bertelli}, {Bressan}, {Chiosi}, {Fagotto},
  and {Nasi}}}]{M/H}
\bibinfo{author}{\bibfnamefont{G.}~\bibnamefont{{Bertelli}}},
  \bibinfo{author}{\bibfnamefont{A.}~\bibnamefont{{Bressan}}},
  \bibinfo{author}{\bibfnamefont{C.}~\bibnamefont{{Chiosi}}},
  \bibinfo{author}{\bibfnamefont{F.}~\bibnamefont{{Fagotto}}},
  \bibnamefont{and} \bibinfo{author}{\bibfnamefont{E.}~\bibnamefont{{Nasi}}},
  \bibinfo{journal}{Astronomy and Astrophysics Supplement Series}
  \textbf{\bibinfo{volume}{106}}, \bibinfo{pages}{275} (\bibinfo{year}{1994}).

\bibitem[{\citenamefont{{Cardelli}
  \emph{et~al.}}(1989)\citenamefont{{Cardelli}, {Clayton}, and
  {Mathis}}}]{Cardelli+1989}
\bibinfo{author}{\bibfnamefont{J.~A.} \bibnamefont{{Cardelli}}},
  \bibinfo{author}{\bibfnamefont{G.~C.} \bibnamefont{{Clayton}}},
  \bibnamefont{and} \bibinfo{author}{\bibfnamefont{J.~S.}
  \bibnamefont{{Mathis}}}, \bibinfo{journal}{Astrophysical Journal}
  \textbf{\bibinfo{volume}{345}}, \bibinfo{pages}{245} (\bibinfo{year}{1989}).

\bibitem[{\citenamefont{{O'Donnell}}(1994)}]{O'Donnell1994}
\bibinfo{author}{\bibfnamefont{J.~E.} \bibnamefont{{O'Donnell}}},
  \bibinfo{journal}{Astrophysical Journal} \textbf{\bibinfo{volume}{422}},
  \bibinfo{pages}{158} (\bibinfo{year}{1994}).

\bibitem[{\citenamefont{{Korn} and {Korn}}(1968)}]{Korn&Korn1968}
\bibinfo{author}{\bibfnamefont{G.~A.} \bibnamefont{{Korn}}} \bibnamefont{and}
  \bibinfo{author}{\bibfnamefont{T.~M.} \bibnamefont{{Korn}}},
  \emph{\bibinfo{title}{{Mathematical handbook for scientists and engineers.
  Definitions, theorems, and formulas for reference and review}}}
  (\bibinfo{publisher}{McGraw-Hill Book Company, New York},
  \bibinfo{year}{1968}).

\bibitem[{\citenamefont{{Yakut} \emph{et~al.}}(2025)\citenamefont{{Yakut},
  {Kalomeni}, and {Rappaport}}}]{Yakut+2025}
\bibinfo{author}{\bibfnamefont{K.}~\bibnamefont{{Yakut}}},
  \bibinfo{author}{\bibfnamefont{B.}~\bibnamefont{{Kalomeni}}},
  \bibnamefont{and}
  \bibinfo{author}{\bibfnamefont{S.}~\bibnamefont{{Rappaport}}},
  \bibinfo{journal}{Monthly Notices of the Royal Astronomical Society}
  (\bibinfo{year}{2025}).

\bibitem[{\citenamefont{{Reyes} \emph{et~al.}}(2024)\citenamefont{{Reyes},
  {Stello}, {Hon}, {Trampedach}, {Sandquist}, and {Pinsonneault}}}]{Reyes+2024}
\bibinfo{author}{\bibfnamefont{C.}~\bibnamefont{{Reyes}}},
  \bibinfo{author}{\bibfnamefont{D.}~\bibnamefont{{Stello}}},
  \bibinfo{author}{\bibfnamefont{M.}~\bibnamefont{{Hon}}},
  \bibinfo{author}{\bibfnamefont{R.}~\bibnamefont{{Trampedach}}},
  \bibinfo{author}{\bibfnamefont{E.}~\bibnamefont{{Sandquist}}},
  \bibnamefont{and} \bibinfo{author}{\bibfnamefont{M.~H.}
  \bibnamefont{{Pinsonneault}}}, \bibinfo{journal}{Monthly Notices of the Royal
  Astronomical Society} \textbf{\bibinfo{volume}{532}}, \bibinfo{pages}{2860}
  (\bibinfo{year}{2024}).

\bibitem[{\citenamefont{{Pang} \emph{et~al.}}(2023)\citenamefont{{Pang},
  {Wang}, {Tang}, {Rui}, {Bai}, {Li}, {Feng}, {Kouwenhoven}, {Chen}, and
  {Chuang}}}]{Pang+2023}
\bibinfo{author}{\bibfnamefont{X.}~\bibnamefont{{Pang}}},
  \bibinfo{author}{\bibfnamefont{Y.}~\bibnamefont{{Wang}}},
  \bibinfo{author}{\bibfnamefont{S.-Y.} \bibnamefont{{Tang}}},
  \bibinfo{author}{\bibfnamefont{Y.}~\bibnamefont{{Rui}}},
  \bibinfo{author}{\bibfnamefont{J.}~\bibnamefont{{Bai}}},
  \bibinfo{author}{\bibfnamefont{C.}~\bibnamefont{{Li}}},
  \bibinfo{author}{\bibfnamefont{F.}~\bibnamefont{{Feng}}},
  \bibinfo{author}{\bibfnamefont{M.~B.~N.} \bibnamefont{{Kouwenhoven}}},
  \bibinfo{author}{\bibfnamefont{W.-P.} \bibnamefont{{Chen}}},
  \bibnamefont{and} \bibinfo{author}{\bibfnamefont{R.-j.}
  \bibnamefont{{Chuang}}}, \bibinfo{journal}{Astronomical Journal}
  \textbf{\bibinfo{volume}{166}}, \bibinfo{eid}{110} (\bibinfo{year}{2023}).

\bibitem[{\citenamefont{{Mikhnevich}
  \emph{et~al.}}(2024)\citenamefont{{Mikhnevich}, {Plotnikova}, {Seleznev}, and
  {Carraro}}}]{Varya_VAK2024}
\bibinfo{author}{\bibfnamefont{V.}~\bibnamefont{{Mikhnevich}}},
  \bibinfo{author}{\bibfnamefont{A.}~\bibnamefont{{Plotnikova}}},
  \bibinfo{author}{\bibfnamefont{A.}~\bibnamefont{{Seleznev}}},
  \bibnamefont{and}
  \bibinfo{author}{\bibfnamefont{G.}~\bibnamefont{{Carraro}}}, in
  \emph{\bibinfo{booktitle}{Modern Astronomy: From the Early Universe to
  Exoplanets and Black Holes (VAK2024}} (\bibinfo{year}{2024}), pp.
  \bibinfo{pages}{443--449}.

\bibitem[{\citenamefont{{Mikhnevich} and {Seleznev}}(2024)}]{WD_bin}
\bibinfo{author}{\bibfnamefont{V.~O.} \bibnamefont{{Mikhnevich}}}
  \bibnamefont{and} \bibinfo{author}{\bibfnamefont{A.~F.}
  \bibnamefont{{Seleznev}}}, \bibinfo{journal}{Astronomy Reports}
  \textbf{\bibinfo{volume}{68}}, \bibinfo{pages}{121} (\bibinfo{year}{2024}).

\bibitem[{\citenamefont{{Cramer}}(1946)}]{Cramer}
\bibinfo{author}{\bibfnamefont{H.}~\bibnamefont{{Cramer}}},
  \emph{\bibinfo{title}{{Mathematical Methods of Statistics}}}
  (\bibinfo{publisher}{Princeton Univ. Press, Princeton, NJ},
  \bibinfo{year}{1946}).

\bibitem[{\citenamefont{{Danilov} and
  {Putkov}}(2012)}]{Danilov_Putkov_2012ARep}
\bibinfo{author}{\bibfnamefont{V.~M.} \bibnamefont{{Danilov}}}
  \bibnamefont{and} \bibinfo{author}{\bibfnamefont{S.~I.}
  \bibnamefont{{Putkov}}}, \bibinfo{journal}{Astronomy Reports}
  \textbf{\bibinfo{volume}{56}}, \bibinfo{pages}{609} (\bibinfo{year}{2012}).

\end{thebibliography}

\end{document}